\documentclass[pra,superscriptaddress, twocolumn]{revtex4}

\usepackage{amsmath}
\usepackage{graphicx}
\usepackage{dsfont}
\usepackage{epstopdf}
\usepackage{amsfonts}
\usepackage{color}
\usepackage{hyperref}
\usepackage{bm}

\definecolor{darkblue}{rgb}{0,0,0.75}
\definecolor{darkred}{rgb}{0.6,0,0}
\definecolor{dg}{rgb}{0,0.3,0}

\hypersetup{
 colorlinks=true,
 linkcolor=darkblue,
 citecolor = darkblue
}

\begin{document}

\title{Nonlinear Atomic Force Microscopy:\\ 
Squeezing and Skewness of Micro-Mechanical Oscillators interacting
with a Surface
}

\author{Karl-Peter Marzlin}
\affiliation{Department of Physics, St. Francis Xavier University, Antigonish, Nova Scotia B2G 2W5, Canada}
\author{Bryan Canam}
\affiliation{Department of Physics, St. Francis Xavier University,
  Antigonish, Nova Scotia B2G 2W5, Canada}
\author{Nisha Rani Agarwal}
\affiliation{Nano-Imaging and Spectroscopy Laboratory,
Faculty of Science, University of Ontario Institute of Technology, 
2000 Simcoe Street North, Oshawa, ON L1G0C5, Canada}

\begin{abstract}
We propose a two-frequency driving scheme in dynamic atomic force microscopy
that maximizes the interaction time between tip and sample. 
Using a stochastic description of the cantilever dynamics, 
we predict large classical squeezing
and a small amount of skewness of the tip's
phase-space probability distribution.
Strong position squeezing will require close contact between tip and
surface, while momentum squeezing would also be possible in the van
der Waals region of the tip-surface force.
Employing a generalized Caldeira-Leggett model,
we predict that surface-dependent dissipative forces
may be the dominant source of quantum effects and 
propose a procedure to isolate quantum effects
from thermal fluctuations.
\end{abstract}

\pacs{42.50.Nn,52.25.Os,52.40.Db}


\maketitle


\section{Introduction}

Surfaces and interfaces, i.e., the outermost layer of atoms
of a material,
define how it interacts with its surroundings \cite{Lueth2015}. It is therefore
critically important to study the physical and chemical properties of surfaces.
For more than three decades,
atomic force microscopy (AFM) \cite{PhysRevLett.56.930,Gavara2017} has
been an indispensable tool to explore the topography of surfaces
and their electrical, magnetic and elastic properties \cite{Magonov2008}.
It has been applied to study a plethora of phenomena 
at the nano-scale, including the measurement
of forces acting on individual molecules \cite{Hinterdorfer2006},
the study of biological samples \cite{Kreplak2016}, and 
nanoparticles \cite{Pyrgiotakis2014}.
Conventional AFM consists of a cantilever with a
nanometer-sized tip that interacts with the sample
while being dragged over its surface (contact mode 
\cite{PhysRevLett.56.930}). The properties
of the sample, and in particular the force between tip and sample, are
measured by monitoring the motion of the cantilever.
In order to minimize the damage to the sample, the cantilever can be
made to rapidly tap over the sample surface while scanning. 
The distance to the surface is
manipulated by applying an oscillating, 
single-frequency piezoelectric force.
This mode is known as dynamic AFM \cite{Garcia2002}.

Our work focuses on intermodulation AFM \cite{Platz2008}, where
the cantilever is driven by a two-frequency force.
We pursue several goals in this work, which
highlight the advantages of dynamic AFM 
and shed light on the conditions and parameter settings required for it.
First, we provide a theoretical
description of the cantilever dynamics within the context of
statistical mechanics. This enables us to study tip
fluctuations, opening a path to monitor additional
properties of the tip-surface interaction. We focus in particular on
squeezing of thermal fluctuations, which could increase the resolution of
AFM. Squeezing in AFM has been studied before 
within a quantum-mechanical model by Passian and Siopsis 
\cite{PhysRevA.94.023812, Passian2017}.
Our second goal is to find a driving scheme that enhances the
generation of squeezing. Since this essentially amounts to maximizing
the effect of the tip-surface interaction, the principle of this
scheme -- maximizing the interaction time -- can be applied to the
measurement of other parameters as well. 
The generation of squeezing in AFM can be compared
to classical nonlinear optics, where a
coupling between different harmonics may trigger
phase squeezing through the
Kerr effect \cite{MandelWolf}.

Our third goal is to explore under which circumstances quantum effects
may be measurable in dynamic AFM. There is a growing interest in 
quantum effects with
micromechanical resonators
\cite{Aspelmeyer2014Book,Bowen2016Book,Riedinger2018}
or even larger objects \cite{MercierdeLepinay2021,Yu2020}.
A mature experimental design such as AFM may help to progress this
field. In normal operations, AFM works well within the classical regime.
For this reason, we have used classical statistical
mechanics for most of our results. However, we also
describe the cantilever as an
open quantum system \cite{GardinerZollerQuantumNoise} to
distinguish quantum and classical dynamics. Of particular interest is the
role of dissipative surface forces, which appear to be particularly
suitable to generate quantum effects.

The paper is organized as follows. In Sec.~\ref{sec:classicalPicture},
we present the driving scheme to maximize the interaction time.
Secs.~\ref{sec:dynamicalEqns} and \ref{sec:skew} are devoted to the
description of the cantilever as a statistical system. In
Sec.~\ref{sec:results}, we discuss our predictions for
the generation of squeezing and skewness.
We expound on the
physics behind squeezing generation in AFM
and the influence of quantum fluctuations in
Sec.~\ref{sec:discussion}. This is followed by a conclusion
\ref{sec:conclusions}.
Several appendices contain the details of our theoretical methods.

\section{Driving scheme to maximize surface interaction}\label{sec:classicalPicture}
Cantilever and tip of AFM form a complex mechanical system that may
include bending motion and torsion \cite{Passian2017}, but a simple
harmonic oscillator model often suffices and will be used here.
The tip is described in phase space 
as a point particle with position $x(t)$ and
momentum $p(t)$ and its dynamics is governed
by Newton's second law,
\begin{align} 
  \dot{p} &= -k x-\gamma_Q  p +F_\text{dr}(t) + F_\text{sf} +
F_\text{dis}.
\end{align} 
Here, $x$ denotes the position of the tip above the sample surface.
The tip is in the engaged position with the sample
for $x\approx -h$. Position $x=0$ corresponds to the
equilibrium position of the free tip, and $k$ denotes the spring constant
of the cantilever. Rate $\gamma_Q$ describes internal mechanical
losses of the cantilever. $F_\text{dr} (t)$
denotes a driving force that puts the cantilever in motion. The
tip-surface interaction is decomposed into a conservative part 
$F_\text{sf}$ and a dissipative part $F_\text{dis}$. The dissipative
part arises from the deformation of the sample surface due to the
interaction with the tip. It can be modelled in different ways, for
instance through 
Kelvin-Voigt viscoelastic dissipation \cite{Benaglia2019},
a hysteretic force \cite{Johnson1971,Dankowicz2008},
a convolution integral \cite{Lee1960}, or a
retarded response to the sample \cite{Ting1966}.
In this
paper, we follow Ref.~\cite{Platz2013} and model the dissipate surface
force by a position-dependent drag force, $F_\text{dis}=-\gamma(x) p$.

\begin{figure}
\begin{center}
\includegraphics[width=7cm]{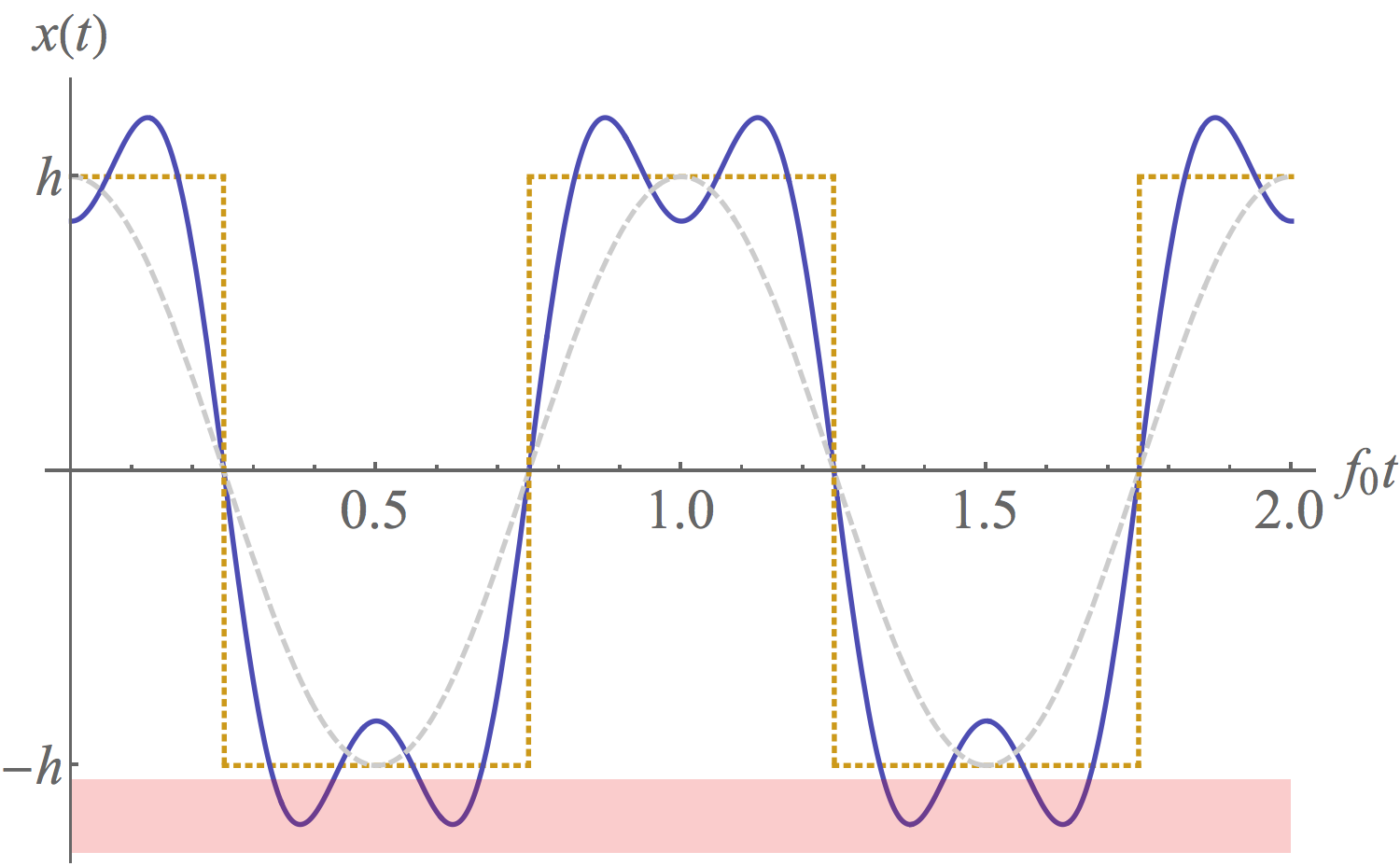}
\caption{\label{fig:drivingScheme} Proposed driving scheme. For
 a periodic motion, the tip would remain close to the
  sample (red rectangle near $x=-h$) if the motion would take the form
  of a square wave (orange
  dotted curve). In single-frequency AFM (gray, dashed), the tip spends little
  time near the sample, but the suggested two-frequency driving scheme
(blue, solid) approximates a square wave.}
\end{center}
\end{figure}
In dynamic AFM, the driving force typically takes the form of a
homogeneous force with one or two frequency components. Here,
we consider two angular frequency components $\omega_1, \omega_2$,
\begin{align} 
  F_\text{dr}(t) &= F_1 \sin(\omega_1 t+\phi_1)
   + F_2 \sin(\omega_2 t+\phi_2),
\label{eq:twoFreqDrivingForce}\end{align} 
with $\phi_1,\phi_2$ controlling the relative phase of the force components. 
In absence of the sample, this model corresponds to an elementary
driven damped harmonic oscillator. After a transient period, the
oscillator will settle to a steady-state motion of the form
\begin{align} 
   x_0(t) &= \frac{ F_1}{M \omega_1} \text{Im}
  \left (
  \frac{ e^{i\omega_1 t+i\phi_1}}{Z(\omega_1)}
  \right )+
  \frac{ F_2}{M \omega_2} \text{Im}
  \left (
  \frac{ e^{i\omega_2 t+i\phi_2}}{Z(\omega_2)}
  \right ),
\label{eq:xUnperturbed}\end{align}
where 
 \begin{align} 
  Z(\omega) &= \frac{ \omega_0^2-\omega^2}{\omega} + i \gamma_Q
\label{eq:defImpedance}\end{align} 
is the mechanical impedance of the cantilever.
$M$ denotes the reduced mass
of cantilever and tip, and $\omega_0 = 2\pi f_0 =\sqrt{k/M}$ the
resonance angular frequency.

The principle of the proposed driving scheme is explained in
Fig.~\ref{fig:drivingScheme}. In dynamic AFM, the driving force
induces a periodic motion of the cantilever. If the force could be
arbitrarily strong, the interaction time with the surface would be
maximized for a square wave motion, where the tip spends half the
time period close to the surface during each cycle. In a two-frequency driving scheme,
one can maximize the time the tip spends near the surface if the two
frequency components of the motion correspond to the first two Fourier
coefficients of a square wave. Eq.~(\ref{eq:xUnperturbed}) implies
that this can be achieved if we pick the driving frequencies as
$\omega_2=3\omega_1$, and the force amplitudes
so that they produce the Fourier coefficients, $\frac{ F_1}{M \omega_1}
\text{Im}e^{i\phi_1} Z^{-1}(\omega_1)=\frac{4}{\pi}h$ and
$\frac{ F_2}{M \omega_2}
\text{Im}e^{i\phi_2} Z^{-1}(\omega_2)=-\frac{4}{3\pi}h$. 
For the choice $\omega_1=\omega_0$,
$Z(\omega_0)$ is imaginary and $Z(3\omega_0)$ is a negative real
number for $\gamma_Q\ll \omega_0$. The phases then can be chosen as
$\phi_1=0$ and $\phi_2=\frac{ \pi}{2}$. However, we will see below
that the optimal choice of phases will depend on the surface interaction when
the tip interacts with the sample.

\section{Dynamical equations for probability moments}\label{sec:dynamicalEqns}
A real tip is not a point particle, and its motion is generally 
subject to thermal or quantum fluctuations, which require
a probabilistic description. The position of the tip is then replaced
by the mean position $\bar{x}(t) = \langle x \rangle $. Here, angle
brackets denote averaging with respect to a probability distribution,
which can be of classical or quantum nature. Variances are described
through mean values of quadratic expressions, such as 
$\Delta x^2 = \langle (x-\bar{x})^2 \rangle $. More generally, one can
characterize a probability distribution through 
its moments \cite{SpanosProbability},
\begin{align} 
  \Delta_{n,m} &= \frac{ 1}{2} 
   \left \langle   (x-\bar{x})^n (p-\bar{p})^m +
                 (p-\bar{p})^m (x-\bar{x})^n 
   \right \rangle .
\label{def:DeltanmMainPart}\end{align} 
In a classical description, the ordering of position and momentum
terms is irrelevant, but we keep a symmetric ordering so that our
formalism can be used for a quantum description as well.
Second-order moments describe position variance  $\Delta x^2 =\Delta_{2,0}$,
momentum variance $\Delta p^2 =\Delta_{0,2}$, and cross-correlation
$\Delta_{1,1}$. 
Higher order moments describe a non-Gaussian 
structure of the distribution. For instance, $\Delta_{3,0}$ describes
the skewness in position, which is a measure for how much the maximum
of the distribution differs from mean value $\bar{x}$ (see section \ref{sec:skew}). 

The driving scheme introduced above has the potential to induce a
large amount of squeezing and a small amount of skewness in the
probability distribution. Squeezing refers to the reduction of the
variance of one observable, say $\Delta x$, at the expense of
increasing the variance of its conjugate variable $\Delta p$. It is
most often considered in the quantum description of light, and it is
sometimes assumed that the coupling of stretching $\Delta p$ while
squeezing $\Delta x$ is a consequence of Heisenberg's uncertainty 
principle. However, squeezing also occurs in classical probability
distributions, where the coupling of stretching and squeezing is a
consequence of the conservation of phase-space volume for conservative
forces because of Liouville's theorem \cite{Liouville1838}.

For a point particle, the dynamical equations take the form
$\dot{p}=F(x)$, but for an object that is described through a
probability distribution, we have
$\dot{\bar{p}} = \langle F(x) \rangle \neq F(\bar{x})$. This implies
that the dynamics of mean position and momentum is generally coupled
to higher-order moments. For instance, $F(x) = c x^3$ would couple
mean momentum $\bar{p}$ to skewness $\Delta_{3,0}$.

For a force that is not given by a power law, the dynamical equation
$\dot{\bar{p}} = \langle F(x) \rangle $, and the equivalent equations
for all moments, will not generate a closed set of equations for mean
values, variances, and skewness (see App.~\ref{sec:FokkerPlanck}
and \ref{sec:basics}). 
To overcome this problem, we 
make a {\em localization approximation}: we assume that the
probability distribution of the oscillator is so narrow that the force
varies little over the width of the distribution. In this case, the
mean force can be expanded as a Taylor series around the mean
position,
\begin{align} 
   \langle F(x) \rangle  &\approx
    \sum_{n=1}^{n_\text{max}}
     \frac{ F^{(n)}(\bar{x})}{n!}
    \langle (x-\bar{x})^n \rangle 
 =  \sum_{n=1}^{n_\text{max}}
     \frac{ F^{(n)}(\bar{x})}{n!} \Delta_{n,0}.
\label{eq:locApproxMainPart}\end{align} 

We have used the localization approximation to derive a coupled set of
equations for mean position $\bar{x}$  and momentum $\bar{p}$, as well as all moments $\Delta_{n,m} $ up to
third order, $(n,m) \in \{ (2,0),(1,1),(0,2),(3,0),(2,1),(1,2),(0,3)\}  $. 
The result is rather lengthy and given by
Eqs.~(\ref{eq:dglX})-(\ref{eq:dD03dt}).
As an example, the dynamical equation for mean momentum 
takes the form
\begin{align} 
     \dot{\bar{p}} &= F 
    -(\gamma_Q +\gamma )\bar{p} 
    + \frac{ \Delta_{2,0}}{2}  \left (  F''
     - \bar{p}   \gamma '' \right )
   - \gamma ' \Delta_{1,1}
\nonumber \\ &\hspace{4mm}
    + \frac{ 1}{6} \Delta_{3,0} \left (
     F^{(3)}  - \bar{p} \gamma^{(3)}
   \right ) 
  - \frac{1}{2}  \gamma ''\Delta_{2,1} .
\label{eq:pbarDot}\end{align} 
In this expression, functions $F$ and $\gamma$ depend on the mean
position, e.g., $F=F(\bar{x})$. We have derived these equations from a
classical stochastic theory (Fokker-Planck equation, see App.~\ref{sec:FokkerPlanck}) and from a
Lindblad-type master equation for open quantum systems (App.~\ref{sec:basics}). 
The results differ by genuine quantum terms, which are proportional to $\hbar^2$
and are highlighted in blue in
Eqs.~(\ref{eq:dglX})-(\ref{eq:dD03dt}). In addition, there is a
technical difference between both derivations that only concerns the
position-dependent dissipative surface force. This difference is
discussed in App.~\ref{sec:basics}.

\section{Tip fluctuations: squeezing and skewness}\label{sec:skew}
Suppose we know the moments $\Delta_{n,m}$, either by measuring it or
by solving the dynamical equations. Can we find the probability
distribution that describes the stochastic state of the oscillator?
The answer is yes, and follows from the reconstruction theorem in
quantum physics \cite{StreaterWightman}, for instance. In
App.~\ref{sec:reconstruction}, we derive an approximate
expression for a classical phase-space probability distribution
$\rho(\boldsymbol{r})$ in terms of moments, where bold symbols
$ \boldsymbol{r}=(x,p)$ 
denote phase-space vectors. A multivariate skew-normal probability distribution
$\rho(\boldsymbol{r})$ can be expressed as the product of a Gaussian
distribution $\rho_0(\boldsymbol{r}) $ and a second function $\Phi$
\cite{Azzalini1996}, 
\begin{align} 
   \rho(\boldsymbol{r}) &= \rho_0(\boldsymbol{r}) \Phi(\boldsymbol{r})
\label{eq:rhoReconstructed}\\
  \rho_0(\boldsymbol{r}) &= \frac{1}{2\pi |C|^{\frac{ 1}{2}}}
   e^{-\frac{ 1}{2} 
   (\boldsymbol{r}-\bar{\boldsymbol{r}})^T\cdot C^{-1} \cdot (\boldsymbol{r}-\bar{\boldsymbol{r}})
   }
\label{eq:GaussianMultiVar} \\
   C &= \left ( \begin{array}{cc}
          \Delta_{2,0}   &  \Delta_{1,1}\\   \Delta_{1,1}&  \Delta_{0,2}
           \end{array} \right ) .
\label{eq:corrMat} \end{align} 
The eigenvalues of correlation matrix $C$ are denoted by
$\sigma_1^2$ and $\sigma_2^2$, where $\sigma_1$ corresponds to the
larger and $\sigma_2$ to the smaller phase-space variance. If the
eigenvectors of $C$ are aligned with position and momentum, position
squeezing corresponds to $\Delta x = \sigma_2$ and $\Delta p=\sigma_1>\sigma_2$
\footnote{In all our calculations, we have expressed $x$ in units of
  the harmonic oscillator ground state width
  $L=\sqrt{\hbar/(M\omega_0)}$ and momentum in units of $\hbar/L$.}.
The Gaussian factor $\rho_0(\boldsymbol{r}) $ would then have an ellipsoid
form that is stretched along the $p$-axis and squeezed along the
$x$-axis. In general, however, squeezing can occur along any direction
in phase space.

If only information about second-order moments is available, the
Gaussian part is all that can be known about $\rho$. Knowledge about
third-order moments enables us to find
the following expansion of $\Phi(\boldsymbol{r})$ around the mean
position $\bar{\boldsymbol{r}} =(\bar{x},\bar{p})$, 
\begin{align} 
   \Phi(\boldsymbol{r}) &= 1 + \boldsymbol{S}\cdot \boldsymbol{R} 
  + \frac{ 1}{6} \sum_{i,j,k}\mu_{ijk} R_i R_j R_k 
   + {\cal O}(\boldsymbol{R}^4)
\label{eq:PhiResult}\\
  S_i &= -\frac{ 1}{2} \sum_{j,k} \mu_{ijk}( C^{-1})_{jk}
\label{eq:shiftVector}\\
  \boldsymbol{R} &= C^{-1}\cdot (\boldsymbol{r}-\bar{\boldsymbol{r}}),
\end{align} 
where $\mu_{111}=\Delta_{30}$, $\mu_{222}=\Delta_{03}$, and
$\mu_{112}=\mu_{121}=\mu_{211}=\Delta_{21}$, as well as
$\mu_{221}=\mu_{212}=\mu_{122}=\Delta_{12}$.
Vector $\boldsymbol{S}$ corresponds to the shift of the maximum of
$\rho$ relative to the mean position $\bar{\boldsymbol{r}}$, as long as
$|\boldsymbol{S}|$ is much smaller than the variances of $\rho$.
The third-order term (the triple sum) generates a roughly triangular 
distortion of the Gaussian profile. 

In the following, we will solve the dynamical equations of motion and
use the reconstructed classical probability distribution to
visualize the effect of the surface force on variances.

\section{Results}\label{sec:results}
To analyze the evolution of the probability distribution
$\rho(\boldsymbol{r})$, we have solved the dynamical equations 
(\ref{eq:dglX})-(\ref{eq:dD03dt}) in two different ways. 
A numerical
solution was found using the software package Mathematica, with
details given in App.~\ref{sec:numDetails}. In
addition, we evaluated the effect of the surface forces using
first-order perturbation theory, with details provided
in App.~\ref{sec:drivenPT}.

The results of both methods indicate that a classical description
is perfectly adequate for typical AFM oscillators. In the discussion below,
we will describe under which circumstances quantum effects may become
relevant, and how one can isolate surface-induced quantum effects from
classical surface-induced effects. Furthermore, both methods predict
that the evolution of mean position and momentum is only very weakly
affected by the coupling to probability moments. For this reason, we
will concentrate on discussing squeezing and skewness.

\begin{figure}
\begin{center}
\includegraphics[width=7cm]{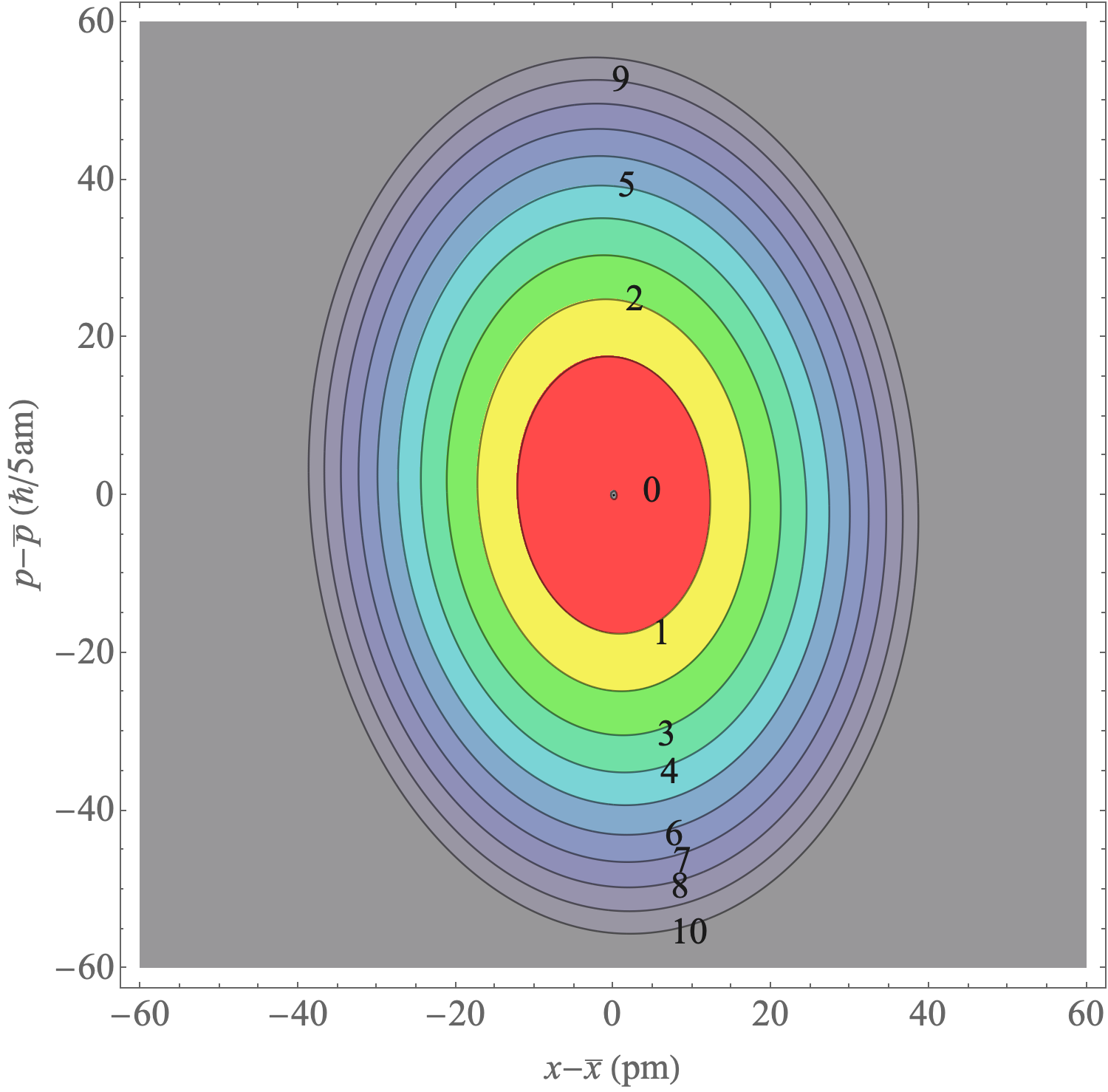}
\caption{\label{fig:rho102} Contour plot of the reconstructed
  classical probability distribution after 102 cycles. A contour
  labeled with $n$ corresponds to a reduction of the probability density by an
amount $e^{-n}$ compared to the maximum value.}
\end{center}
\end{figure}

\begin{figure}
\begin{center}
a)\includegraphics[width=7cm]{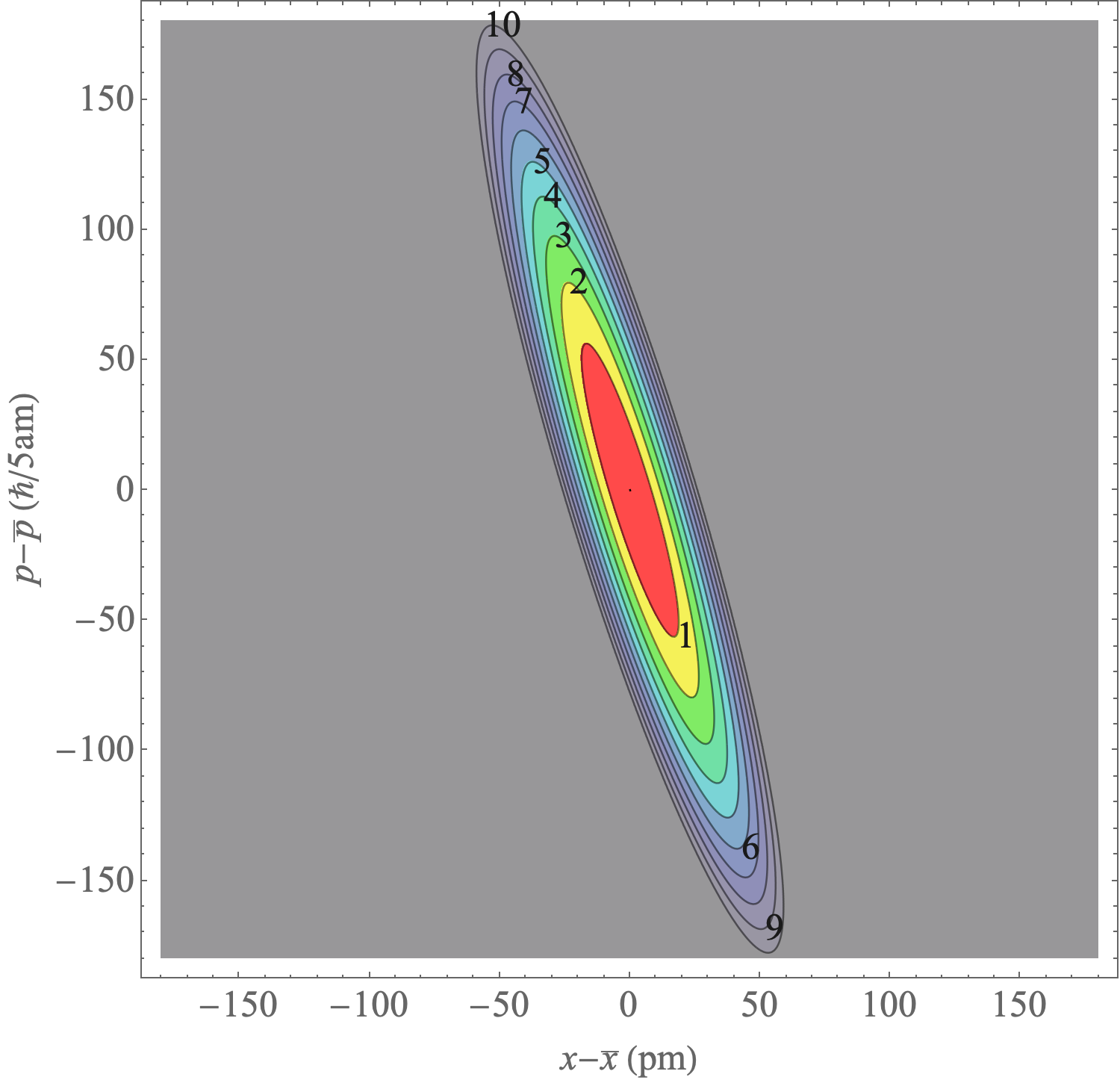}
\\
b) \includegraphics[width=7cm]{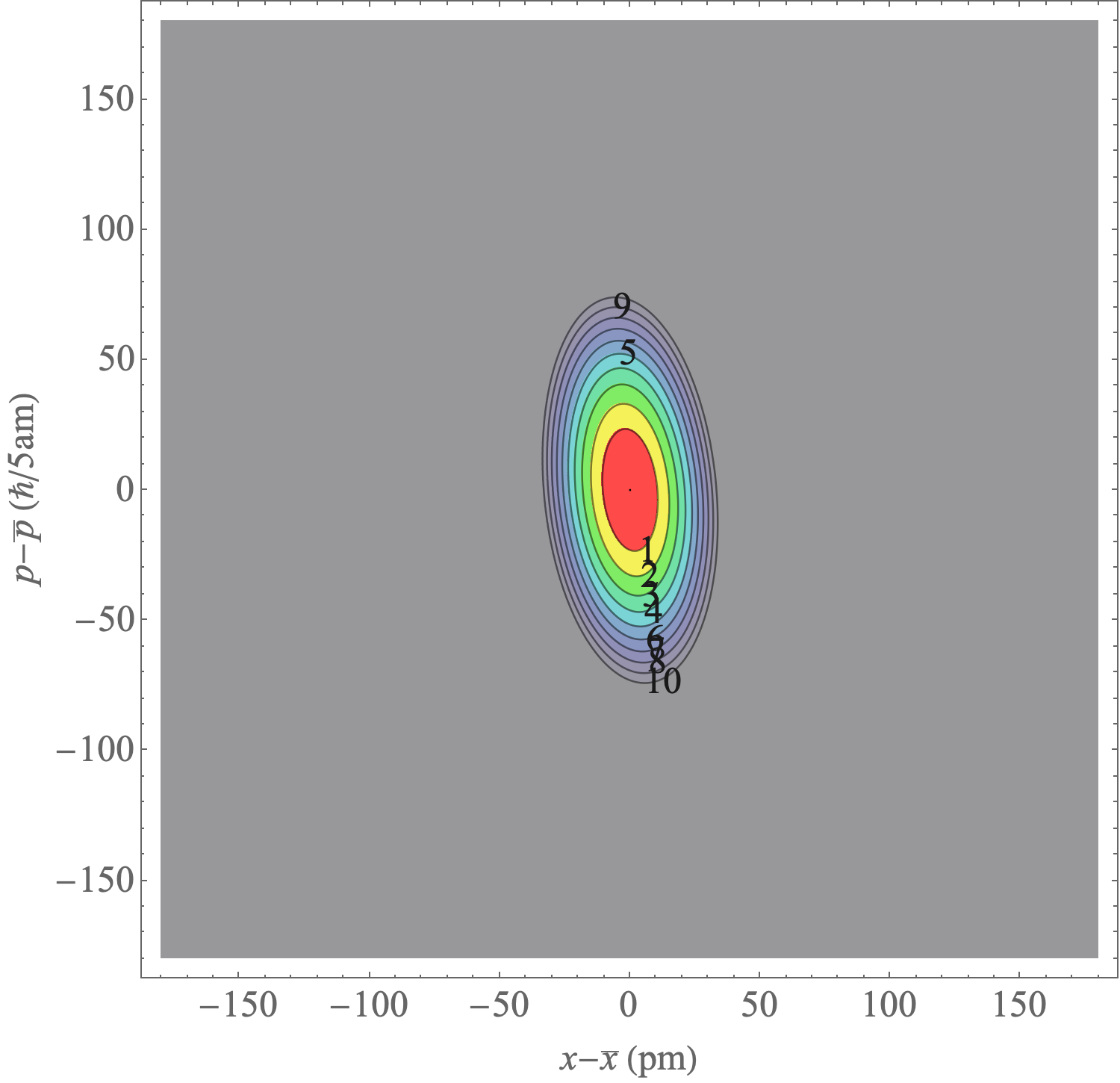}
\\
c) \includegraphics[width=7cm]{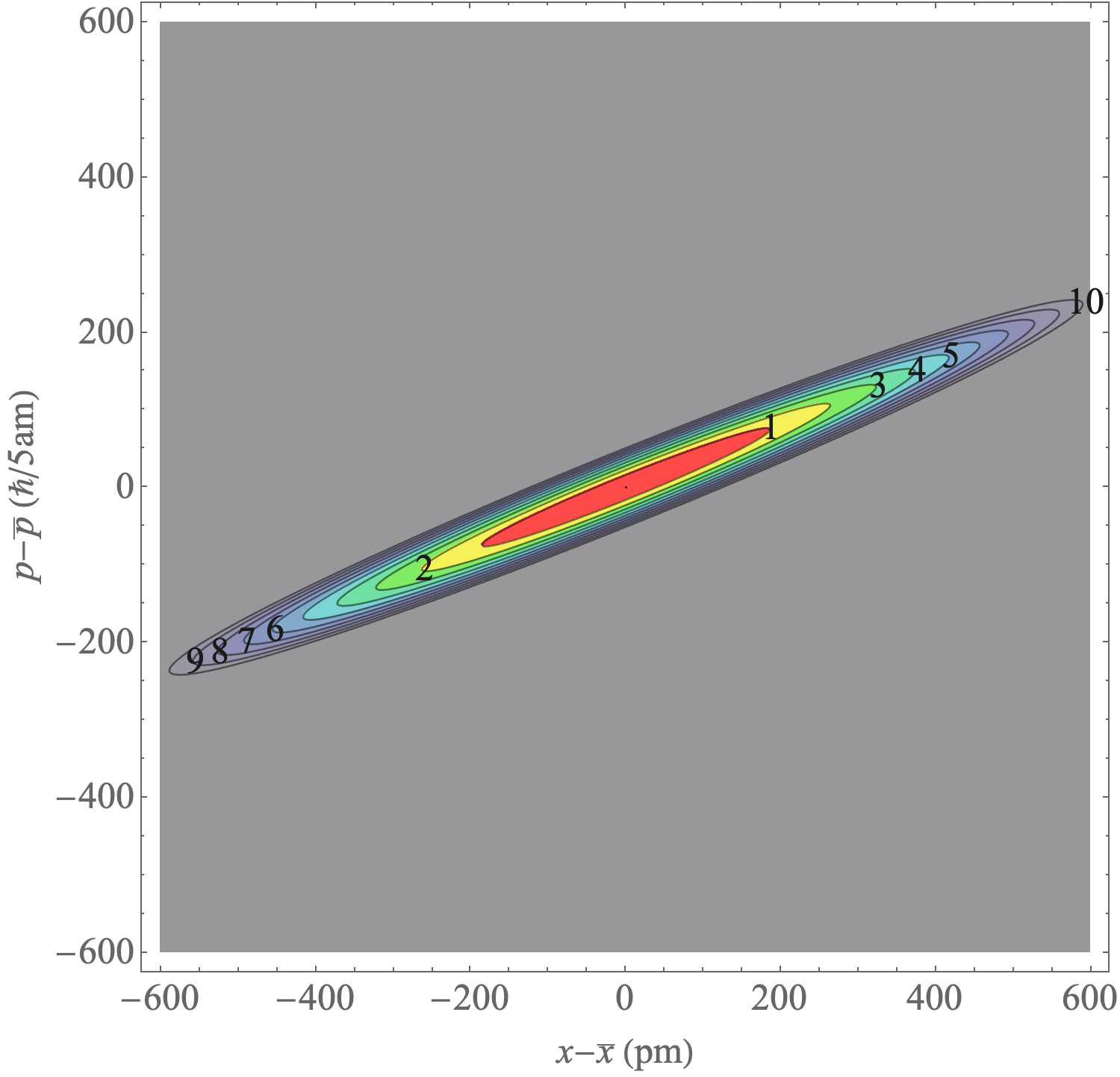}
\caption{\label{fig:rho198} a) Contour plot of the reconstructed
  classical probability distribution after 198 cycles, for
  $\phi_1=\pi$ rad and $\phi_2 = 2.67$ rad.
b) The same for $\phi_1=0$ rad and $\phi_2 = \frac{ \pi}{2}$.
c) The same as a) but with strongly reduced dissipative surface force.}
\end{center}
\end{figure}

In our numerical simulations, we have considered a cantilever with
a resonance frequency of 300 kHz and studied the time evolution for
up to 200 cycles. This is a typical duration for many AFM experiments,
and it is sufficiently long to induce strong squeezing. We have
considered several different cases, including single-frequency on- and
off-resonant driving, as well as two-frequency driving with different
sets of frequencies $\omega_i$ and phase factors $\phi_i$ in 
driving force (\ref{eq:twoFreqDrivingForce}). Figures \ref{fig:rho102}
and \ref{fig:rho198} show the reconstructed probability distribution
at the time when the tip is close to the sample during cycle 102 and
198, respectively \footnote{The momentum units used in the graphs
  may appear unusual. Squeezing is best observed if position and
  momentum are measured in units of $L$ and $\hbar/L$,
  respectively. In this case, position and momentum uncertainties are
  equal, both for the quantized ground state of the oscillator, and
  for a thermal classical distribution. We have used these natural
  squeezing units in our calculations. For the graphs, we have used
  the same rescaling factor for position and momentum to represent
  position in pm. As a consequence, momentum is then measured 
in $\hbar$ over atto meters.}. 
For this specific case, we have used 
$\omega_1=\omega_0, \omega_2=3\omega_0$, as well as $F_1 = 1.21 $ nN
and $F_2=950$ nN. The second force component needs to be much stronger
to compensate for the fact that it drives the cantilever off-resonantly.
The phases of the force were chosen as $\phi_1=\pi$ rad
and $\phi_2 = 2.67$ rad. This choice of phases is rather different
from the phases presented at the end of
Sec.~\ref{sec:classicalPicture} for reasons we will discuss below.

We have chosen to display the reconstructed probability distribution
after 102 cycles, because at this time triangular distortions of the
Gaussian distribution, i.e., the triple sum in
Eq.~(\ref{eq:PhiResult}), are strongest. Since these distortions grow
with the third power of the distance vector $R_i$ from the mean
position of the tip, their effect is more pronounced far away from the
center of the distribution. However, even on the outermost contour 
in Fig.~\ref{fig:rho102}, the deviation from a Gaussian shape only
amounts to about 2\% and is barely visible. Shift vector $\mathbf{S}$
of Eq.~(\ref{eq:shiftVector}), which quantifies skewness,
is negligibly small at this time, about 0.01 pm.

After 198 cycles, strong squeezing can be observed, see
Fig.~\ref{fig:rho198} a). For our choice of parameters, we observe
almost a pure position squeezing (i.e, the ellipse is almost
vertical). Skewness has grown
by a factor of 10 to about 0.15 pm. However, like triangular
distortions, the effect is too small to be visible in our plots. 

\begin{figure}
\begin{center}
a)\includegraphics[width=7cm]{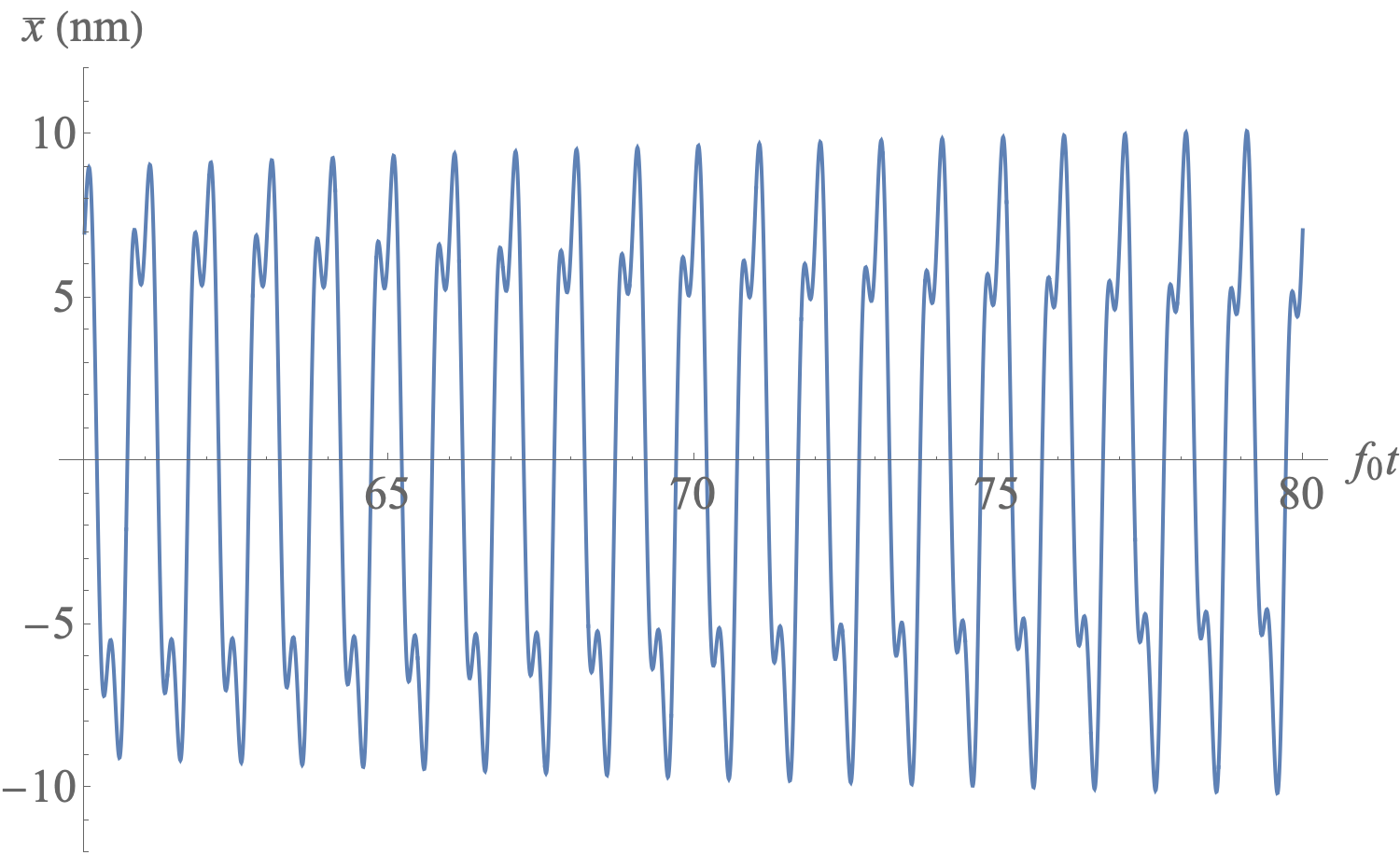}
\caption{\label{fig:xMean} Evolution of the tip's mean position over
  several periods $T=1/f_0$ for a driving force with phase factors
 $\phi_1=\pi$ rad and $\phi_2 = 2.67$ rad.}
\end{center}
\end{figure}

We found that the amount of squeezing critically depends on the choice
of phase factors for the driving force. For instance,
Fig.~\ref{fig:rho198} b) shows the same situation, but with phase
factors chosen as described in Sec.~\ref{sec:classicalPicture}. The
reason for the strong reduction in squeezing is that the 
surface force has a strong influence on the
double-peak structure in Fig.~\ref{fig:drivingScheme}, which is 
key to increasing the tip-surface interaction time. 
In the presence of
the surface, the double peak tends to become asymmetric, so that the
tip is not in close contact with the surface for a longer time
anymore, see Fig.~\ref{fig:xMean}.
The reason for this strong influence is that the relative sign of the
two Fourier components of the approximate square wave in 
Fig.~\ref{fig:drivingScheme} matters a lot for its overall shape. In
the figure, the two Fourier components have opposite signs and thus
cancel each other out close to the central peak. However,
if they have the same sign, they add up to a more pronounced single
peak. 

In Fig.~\ref{fig:xMean}, one can see that the double-peak structure
varies over time. The reason is that the relative phase of the two
frequency components is affected by the surface force over time. The
double peak
therefore varies between an approximate square wave and a single peak.
By adjusting the phase factors of the two driving force components,
one can mitigate this effect to some extent and thus optimize
squeezing. Another possibility would be to adjust the driving
frequencies slightly, to compensate for the temporal variation of the
double peak.

Finally, we have studied the effect of the dissipative surface force
on squeezing by comparing the result of our full simulation 
Fig.~\ref{fig:rho198} a) to a simulation in which the dissipate
surface force is reduced by a factor of $10^{-3}$. The result is shown in
Fig.~\ref{fig:rho198} c) and indicates that $F_\text{dis}$ both
suppresses squeezing and has a strong influence on the axis of the
ellipse. We generally found that, without $F_\text{dis}$, one almost
always obtains momentum squeezing rather than position squeezing
when the tip is close to the sample.

\section{Discussion}\label{sec:discussion}
Our results show several general trends, which are discussed in
this section.

The role of the {\em dissipative surface force} is two-fold: it reduces
the amount of squeezing, and it rotates the squeezing axis so that one
can have (mostly) position squeezing instead of momentum
squeezing.
Since the fluctuation-dissipation theorem
\cite{PhysRev.83.34,Kubo1966} predicts an
increase of fluctuations if a dissipative force acts on a system,
it is natural that squeezing is reduced;
reducing variances simply becomes more difficult.
In our dynamical equations, fluctuations are
described by terms proportional to $p_\text{th}$ in
Eq.~(\ref{dgl:dp2}). A dissipative force can change the orientation of
the squeezing axis through its momentum dependence, which enables it
to counteract features of the conservative surface force that will be
discussed in the next paragraph.

\begin{figure}
\begin{center}
\includegraphics[width=8cm]{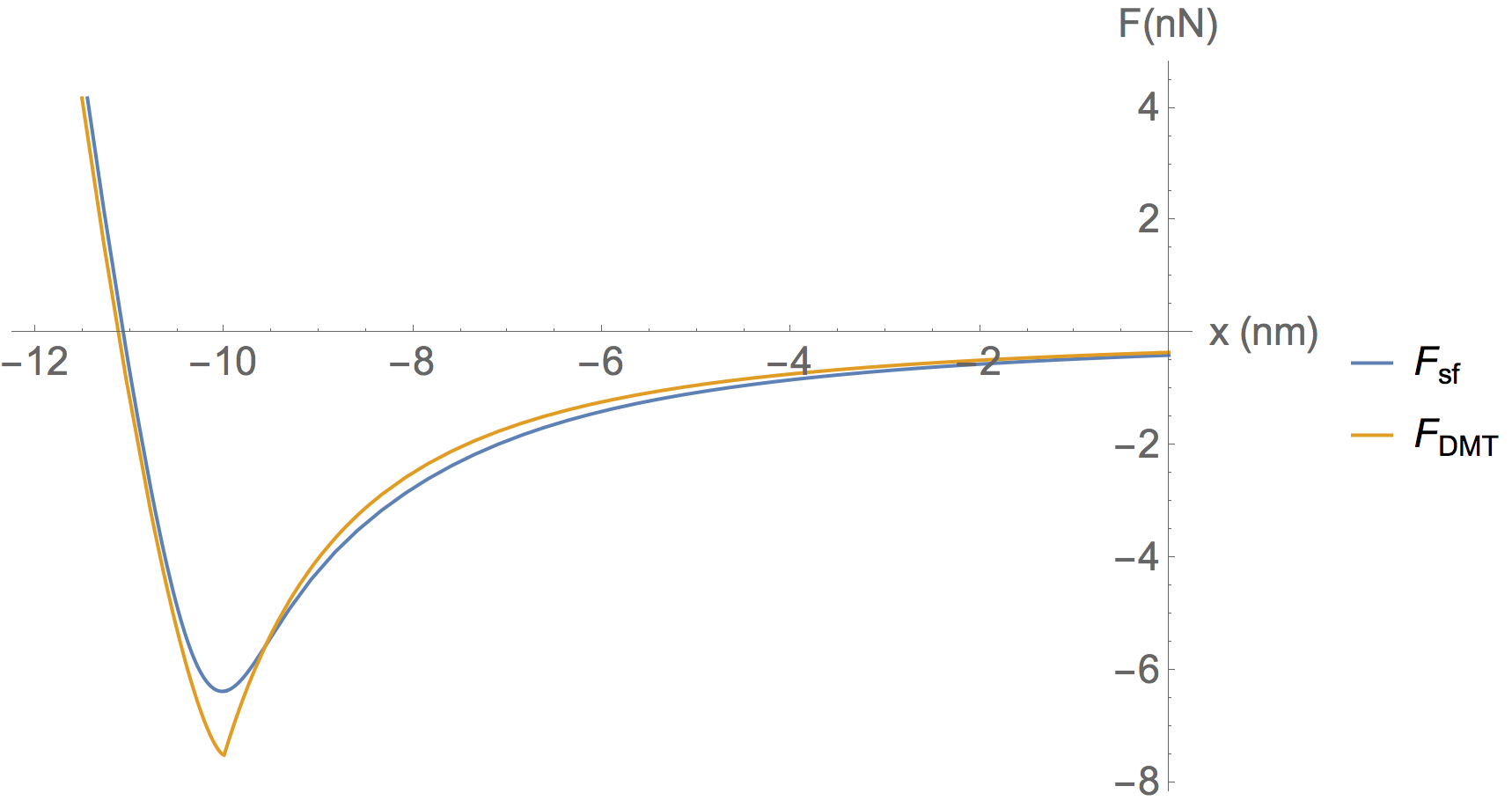}
\caption{\label{fig:forcePlots} Conservative surface force
 $F_\text{sf}$ as a function of cantilever position $x$. The blue curve
shows the modified DMT model (\ref{eq:WDMTmodelSmooth}) used in this work.
The orange curve displays the original DMT model (\ref{eq:WDMTmodelOriginal}).}
\end{center}
\end{figure}

The {\em conservative surface force} alone tends to create momentum
squeezing. We conjecture that the reason for this is that, for most of
the time during one oscillation cycle, the tip is moving in the long-range van der Waals tail of
the surface force. As can be seen in Fig.~\ref{fig:forcePlots}, this
tail has negative curvature, i.e., the force would pull particles that
are closer to the sample stronger towards the surface than particles
that are further away. Consequently, a phase-space probability distribution
would then tend to be stretched along the spatial axis. Since $F_\text{sf}$ is a
conservative force, Liouville's theorem then implies squeezing in 
momentum direction.

Conversely, a conservative force with positive curvature would tend to
squeeze the spatial direction of a probability distribution
and stretch it in momentum direction. For the DMT force shown in Fig.~\ref{fig:forcePlots}, 
this is the case when the tip is in close contact with the surface
($x<-10 $ nm). 
Furthermore, our numerical results suggest that this may 
also be the 
most effective way to achieve large squeezing. 
The reason is that the coupling between
$\Delta_{1,1}$ and $\Delta_{0,2}$ 
(the first term on the right-hand side of Eq.~(\ref{dgl:dp2})), which
is responsible for the generation of squeezing, is
proportional to the gradient of the force. The gradient 
of the surface force is much
larger in the contact region than in the van der Waals
region, so that a larger amount of squeezing can be achieved.

\begin{figure}
\begin{center}
\includegraphics[width=8cm]{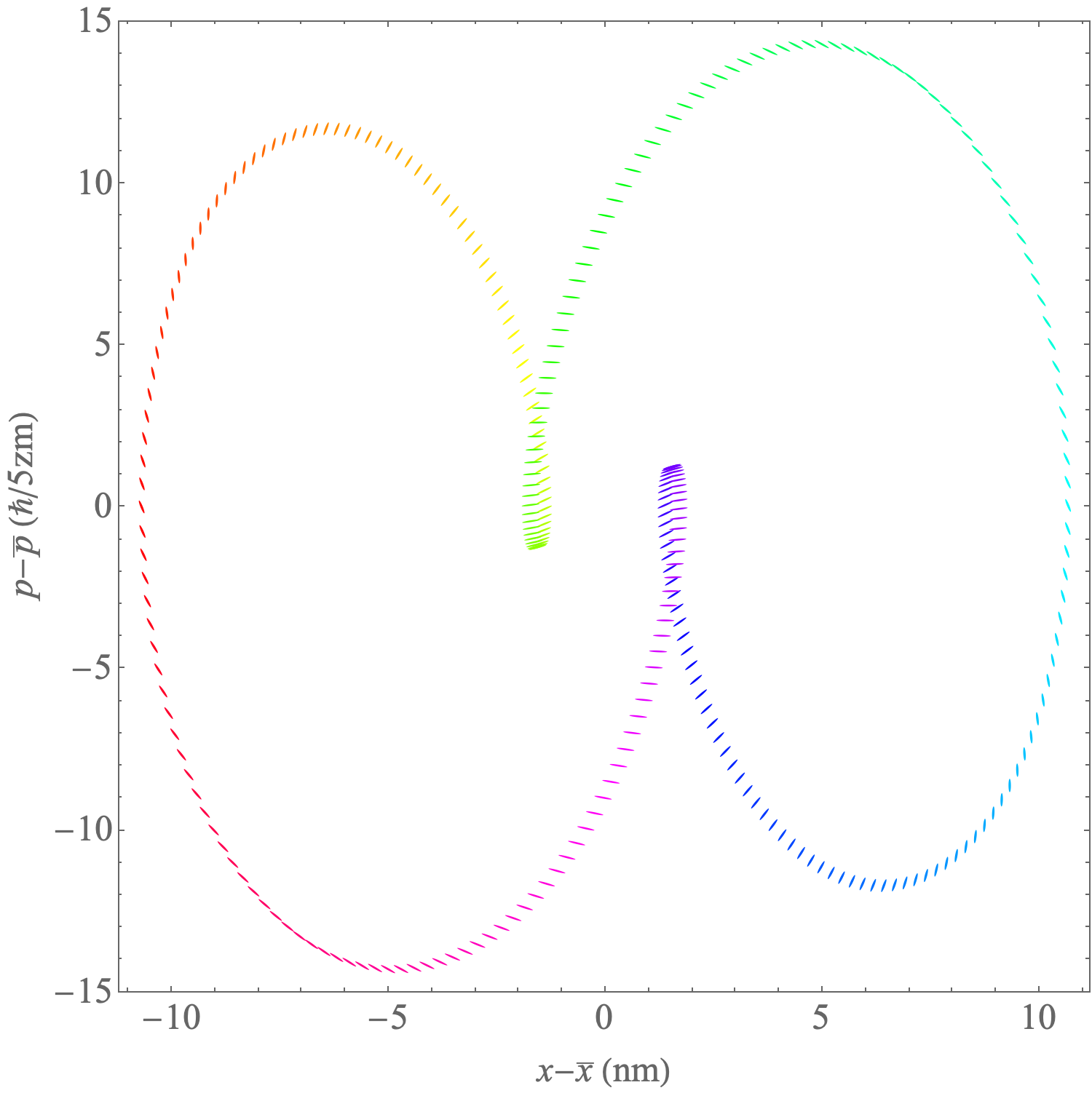}
\caption{\label{fig:fullCycle198} Time evolution of the reconstructed
  probability distribution during the full 198th cycle. Each ellipse
  corresponds to the probability distribution at one particular time
  during the cycle, and is centered around the mean position of the
  tip at that time.
  The size of the ellipse corresponds to the mean variances of the distribution in phase
  space. For instance, the red ellipse in Fig.~\ref{fig:rho198} a)
  corresponds to the leftmost ellipse in this figure. 
For better visibility, the size of each ellipse has been
increased by a factor of 5.}
\end{center}
\end{figure}

An alternative way to achieve position squeezing is the
interplay between a conservative and  a non-conservative
force, such as $F_\text{dis}$.
The details of such an interplay are involved due to the overall
dynamics of the tip during a cycle, which is shown in Fig.~\ref{fig:fullCycle198}. 
For a two-frequency driving force, the tip's mean position follows a
non-circular path. In addition, the axes of the ellipse oscillate at twice the
resonance frequency. The latter effect is known from optical squeezing
and follows from our perturbative treatment in App.~\ref{sec:drivenPT}. 

{\em Quantum effects} are generally negligible in our numerical
simulations. To understand how they could be increased, we
used first-order
perturbation theory to study the influence of a dissipative surface force $F_\text{dis}$ of form  
(\ref{eq:PlatzDissip}). We concentrate on the dissipative part because
it generates terms proportional to $\hbar^2$ in the dynamical equations for
variances $\Delta_{2,0}, \Delta_{1,1},\Delta_{0,2}$, whereas the
conservative surface force only introduces quantum terms for
third-order moments. The first-order correction to the position
variance is given by
\begin{align} 
 \Delta_{20}^{(1)} &= \gamma_0 \int_0^t dt' e^{-\gamma _Q (t-t')} 
   e^{-\frac{h }{x_{\gamma }} 
        \left(1+\cos (\omega _0 t'+\phi_1)\right) }
  \times
\nonumber \\ &\hspace{4mm}
  \Bigg [
  \frac{\left(L^4-16 \Delta x_{\text{th}}^4\right) 
  }{8 \omega_0 x_\gamma^2}
  - \frac{ \gamma _Q }{ \omega _0}
  \frac{ L^4 
   }{8 \Delta x_{\text{th}}^2} 
  \sin \left(2 \omega _0 (t-t')\right)
\nonumber \\ &\hspace{4mm}
-\frac{
   \left(L^4 \left(\Delta x_{\text{th}}^2+x_{\gamma }^2\right)
    -16\Delta x_{\text{th}}^6\right)
   }{8 \Delta x_{\text{th}}^2 x_{\gamma }^2}
  \cos \left(2 \omega _0 (t-t')\right)
  \Bigg ].
\label{eq:ptResult}\end{align} 
This result applies to the case of a single-frequency driving force
($F_2=0$ in Eq.(\ref{eq:twoFreqDrivingForce})). In this case, 
the unperturbed mean position (\ref{eq:xUnperturbed}) varies like
$x_0(t) = h \cos (\omega _0 t+\phi_1)$.

In Eq.~(\ref{eq:ptResult}), all quantum
terms are proportional to $L^4$. Compared to terms that are
proportional to thermal fluctuations
$\Delta x_{\text{th}}^4$, quantum terms are suppressed by a factor 
$L^4/\Delta x_{\text{th}}^4 =(\hbar \omega_0/(k_B T))^2$. 
For current AFM designs and at room temperature, this is only about
$10^{-15}$, but for a 10-fold increased resonance frequency and cooled
to liquid Helium temperatures, this ratio could be increased to $10^{-8}$. While
this is still small, physicists have developed powerful methods to
eliminate the effect of noise from a signal. This includes
spin echoes \cite{PhysRev.80.580},
Doppler-free spectroscopy \cite{PhysRevLett.34.307},
force-insensitivity in atom interferometry \cite{PRA53:312},
and correlation measurements in AFM \cite{Pottier2017}.

Here, we propose a procedure to separate quantum fluctuations
from thermal fluctuations in AFM by adjusting the phase of the driving force.
We start with the observation that, if a term proportional to
$L^4$ appears together with $\Delta x_{\text{th}}^4$ in one algebraic
expression, it is not possible to separate these two terms.
This applies to the first and third term inside the square brackets
in Eq.~(\ref{eq:ptResult}).
Hence, the
only quantum term that can potentially be separated is the term proportional to
$\sin \left(2 \omega _0 (t-t')\right)$.

To isolate this term, we 
first note that the first term in square brackets is constant, while the
other two terms are oscillating at frequency $2\omega_0$. Hence, the
first term can be eliminated through spectral analysis. 

To eliminate the second classical term, which is
proportional to $\cos \left(2 \omega _0 (t-t')\right)$,
we note that the exponential factor in Eq.~(\ref{eq:ptResult})
suppresses the integrand unless the tip is close to the sample, i.e.,
when $\cos (\omega _0 t'+\phi_1)\approx -1$, or
$\omega _0 t'+\phi_1 \approx \pi (2n+1)$, with $n \in \mathds{Z}$.
This is a direct consequence of $F_\text{dis}$ decreasing
exponentially with the distance from the surface.

Let us now for simplicity
assume that we observe the signal periodically at times $t$ that are multiples of
the AFM period, $\omega_0 t = 2\pi m$ with $m \in \mathds{Z}$.
We then have $\cos \left(2 \omega _0 (t-t')\right)
= \cos \left(2 \omega _0 t'\right) $, or when the tip is close to the surface,
$\cos \left(2 \omega _0 (t-t')\right)\approx 
\cos(2\pi(2n+1)-2\phi_1) =\cos(2\phi_1)$.
Hence, if we set $\phi_1 =\frac{ \pi}{4} + n' \frac{ \pi}{2}$, with
$n'\in \mathds{Z}$,
we ensure that $\cos \left(2 \omega _0 (t-t')\right) \approx 0$ when
the tip is close to the surface. As a consequence, the thermal noise
contribution will be strongly suppressed.

The above argument is supported by a numerical evaluation of the
cosine integral as a function of $\phi$ for parameter settings
$Q=400, h=6.7x_\gamma$, and $t=400\pi/\omega_0$. We found that the integral 
then varies like $\cos(2\phi_1)$ and indeed vanishes for specific
choices of the driving force's phase. If the signal is monitored
with period $1/f_0$ at other times $t$, the phase $\phi_1$ of the driving force
can be adjusted so that the thermal signal is suppressed as well.

\section{Conclusion}\label{sec:conclusions}
We have proposed a driving scheme to increase the interaction time between 
tip and sample in intermodulation AFM.
The tip is described as a stochastic system that can exhibit fluctuations. 
We derived a set of coupled dynamical equations for probability
moments, which have some similarity to
the BBGKY hierarchy in statistical mechanics \cite{Cercignani97}.
The solution to these equations enables us to 
reconstruct the tip's phase-space  probability 
distribution.

We use the driving scheme to investigate the generation of
squeezing in the classical phase space of the tip. We predict that
strong position squeezing is possible if the tip is in close contact
with the surface, and if the phases of the two frequency components
in intermodulation AFM are chosen appropriately.
In the weakly interacting van der Waals regime of
the tip-surface interaction, momentum squeezing is predominant. The
dissipative part of the surface force has a strong influence on amount
and orientation of phase-space squeezing.

To distinguish between classical and quantum effects, we
derived the dynamical equations both
using a classical Fokker-Planck equation and
a quantum master equation. We found that
AFM is generally
well described by a classical model. Quantum effects tend to be
much smaller than thermally induced fluctuations, but it is possible
to separate both effects by adjusting the phase of the driving
force in dynamic AFM.

A particularly interesting result is that dissipative surface forces
can introduce quantum dynamics already at the level of variances,
whereas quantum effects induced by a conservative force
are tied to third-order (or higher) probability moments. The derivation
of our dynamical equations in App.~\ref{sec:basics} indicates that the
reason is a particular feature of the dissipative force.
In the model we studied, it is a
position-dependent drag force that depends on two non-commuting
observables: position and momentum.

We conjecture that similar
effect would also occur with other models for dissipative surface
forces. For instance, in the hysteretic JKR model \cite{Johnson1971}, the
surface force changes depending on whether the tip moves towards or
away from the sample; i.e., it depends on position and on
the sign of the tip's
momentum. In a retarded model \cite{Ting1966}, the dissipative
surface force when the tip is at position $x(t)$ depends on an earlier
position $x(t-t_r)$, where $t_r$ is the retardation time of the
force. In Heisenberg picture, $x(t)$ and $x(t-t_r)$ are generally
non-commuting.

By probing the probability distribution of the tip, one can examine
different Fourier components of the surface force, study its
fluctuations,
and enhance specific
effects, similar to methods used in nonlinear optics. Future
work may address the question whether our driving scheme may
also be helpful to measure magnetic surface forces, or if the separation scheme
for quantum effects could be useful to isolate specific classical
effects as well.

\acknowledgements
We would like to thank the Natural Sciences and Engineering Research Council of Canada (NSERC) for financial support. 

\newpage
\begin{widetext}
\begin{appendix}
\section{Classical derivation of dynamical equations based on the Fokker-Planck equation}\label{sec:FokkerPlanck}
For a position-independent drag force $-\gamma_Q p$, the Fokker-Planck
equation \cite{Fokker1914,Planck1917}
for  a classical
phase-space probability distribution $f(x,p,t)$ can be written as
\begin{align} 
  \partial_t f &= - \frac{ p}{M} \partial_x f - \partial_p
                 ((F(x,t)-\gamma_Q p)f)
              +  \frac{g}{2}\partial_p^2 f,
\end{align} 
with $g=2M \gamma_Q k_B T $.
The mean value of a classical observable 
$A(x,p,t)$, where we admit an explicit time dependence, is given by
\begin{align} 
  \langle A \rangle &= \int dx\, dp\, f(x,p,t) \, A(x,p,t) .
\end{align} 
As above, we use the notation $ \bar{x}(t) := \langle x \rangle$ and
$ \bar{p}(t) := \langle p \rangle$. 
The time derivative of a mean value can be expressed as
\begin{align} 
   \frac{ d}{dt}\langle A \rangle &= 
  \langle \partial_t A \rangle 
   +\frac{ 1}{M} \langle p \partial_x A \rangle 
  + \langle (F-\gamma_Q p) \partial_p A \rangle 
  + M \gamma_Q k_B T  \langle \partial_p^2A \rangle .
\label{eq:dMeanA}\end{align} 

To add a position-dependent drag force $-\gamma(x) p$, we start with
the Langevin equation for such a force,
\begin{align} 
   \dot{x} &= \frac{ p}{M}
\\
  \dot{p} &= -\gamma(x) p+ \zeta,
\end{align} 
where $\zeta$ is a noise force. We assume $\delta$-correlated noise,
for which $ \langle \zeta(t) \rangle =0$ and
$\langle \zeta(t)\zeta(t') \rangle = g(t) \delta(t-t')$. The unknown
function
$g(t)$ can be found using the
solution for the momentum,
\begin{align} 
  p(t)&=p(0) e^{-\bar{\Gamma}(t)} 
  +\int_0^t dt'\,
  e^{-\bar{\Gamma}(t)+\bar{\Gamma}(t')} \zeta(t')
\\
  \bar{\Gamma}(t) &= \int_0^t dt' \, \gamma(x(t')).
\end{align} 
The mean kinetic energy of the particle is given by $E=\langle p^2
\rangle /(2M)$. For large times, for which 
$e^{-\bar{\Gamma}(t)} \approx 0$, the
particle should reach its equilibrium energy $E= \frac{ 1}{2} k_B
T$. Hence,
\begin{align} 
  M k_B T &= \int_0^t dt'\,  \int_0^t dt''\,
  e^{-2\bar{\Gamma}(t)+\bar{\Gamma}(t') +\bar{\Gamma}(t'')} 
  \langle \zeta(t') \zeta(t'') \rangle 
\\ &=
 \int_0^t dt'\,
  e^{-2\bar{\Gamma}(t)+2\bar{\Gamma}(t') } g(t')
\\ &=
 \int_0^t d\tau\,
  e^{-2\bar{\Gamma}(t)+2\bar{\Gamma}(t-\tau) } g(t-\tau).  
\end{align} 

If the particle does not move too far during the relaxation process,
we can use the approximation $g(t-\tau)\approx g(t)$
and $\bar{\Gamma}(t-\tau)\approx \bar{\Gamma}(t)-\tau
\gamma(x(t))$. We can also extend the integration interval to infinity.
We then obtain 
\begin{align} 
  M k_B T \approx \frac{ g(t)}{2\gamma(x(t))}.
\end{align} 
This implies that the factor $g$ takes the same form as for the constant drag
force, with $\gamma_Q$ replaced by $\gamma(x(t))$. This is the
approximation we will use below, but we emphasize that this is
model-dependent. For instance, if we use instead the approximation
 $g(t-\tau)\approx g(t)-\tau \dot{g}(t) $, we instead obtain the equation
\begin{align} 
  M k_B T \approx \frac{ g(t)}{2\gamma(x(t))} - \frac{ \dot{g}(t)}{4\gamma^2(x(t))},
\end{align} 
which has the solution
\begin{align} 
  g(t) &\approx \frac{ p_\text{th}^2}{2} 
  e^{
   2\int_0^t \gamma(x(t')) dt'  
  }
  \left \{
   \gamma(0)-2 \int_0^t dt''\,\gamma^2(x(t''))
     e^{
   -2\int_0^{t''} \gamma(x(t')) dt'  
  }
  \right \}.
\end{align} 
Obviously, the precise form of $g(t)$ is difficult to find in a
classical model. However, with the simple approximation, we obtain a
reasonable agreement with the quantum model.
Putting everything
together, we obtain the Fokker-Planck equation
\begin{align} 
 \frac{ d}{dt}\langle A \rangle   &= 
  \langle \partial_t A \rangle 
   +\frac{ 1}{M} \langle p \partial_x A \rangle 
  + \langle (F-(\gamma_Q+\gamma(x)) p) \partial_p A \rangle 
  + M  k_B T
  \langle  (\gamma_Q+\gamma(x))  \partial_p^2A \rangle .
\end{align} 

To derive the classical equations of motion for the correlation
functions (\ref{def:DeltanmMainPart}), we consider observables of the form
$A=(x-\bar{x})^n(p-\bar{p})^m$, so that
\begin{align} 
   \langle \partial_t A \rangle &=-n \dot{\bar{x}} \Delta_{n-1,m} -m
                                  \dot{\bar{p}} \Delta_{n,m-1}
\\ &= -n \frac{ \bar{p}}{M} \Delta_{n-1,m} -m 
   \left\langle F-(\gamma_Q+\gamma(x))p\right \rangle \Delta_{n,m-1}
\\ 
  \langle p \partial_x A \rangle &= n\left (
  \bar{p}\Delta_{n-1,m} +\Delta_{n-1,m+1}
  \right )
\\
   \langle \partial_p^2A \rangle &= m(m-1) \Delta_{n,m-2}.
\end{align} 
We then obtain
\begin{align} 
  \dot{\Delta}_{n,m} &= \frac{n}{M} \Delta_{n-1,m+1} 
   - m \gamma_Q \Delta_{n,m} 
   +m  \left \{ 
   \langle F-\gamma(x)p (x-\bar{x})^n(p-\bar{p})^{m-1} \rangle 
    -\langle F-\gamma(x)p \rangle \Delta_{n,m-1}
  \right \}
\nonumber \\ &\hspace{4mm}
  + M k_B T m(m-1) 
   \left \{ \gamma_Q \Delta_{n,m-2} +
   \langle \gamma(x) (x-\bar{x})^n(p-\bar{p})^{m-2} \rangle 
  \right \}.
\label{eq:FPbeforeLA}\end{align} 

To obtain a closed set of coupled dynamical equations,
we assume that 
uncertainties in position and momentum in the tip motion will remain small, and the tip remains localized.
If the forces acting on the system vary little across the extension of
the probability distribution, we can expand $F(x)$ and $\gamma(x)$ in a Taylor series
around mean position and momentum,
\begin{align} 
   F(x) &\approx
    \sum_{n=1}^{n_\text{max}} \frac{ F^{(n)}(\bar{x})}{n!}
    (x-\bar{x})^n
\label{eq:locApprox}\\ 
     \gamma(x) &\approx
    \sum_{n=1}^{n_\text{max}} \frac{ \gamma^{(n)}(\bar{x})}{n!} 
    (x-\bar{x})^n.
\end{align} 
Applying this approximation to Eq.~(\ref{eq:FPbeforeLA}) leads to
\begin{align} 
  \dot{\Delta}_{n,m} &\approx \frac{n}{M} \Delta_{n-1,m+1} 
   - m \gamma_Q  \Delta_{n,m} 
  + M k_B T m(m-1) 
   \left \{ \gamma_Q \Delta_{n,m-2} +
   \sum_{l=0}^{l_\text{max}} \frac{ \gamma^{(l)}(\bar{x})}{l!} \Delta_{n+l,m-2}
  \right \}
\nonumber \\ &\hspace{4mm}
   +m  \sum_{l=0}^{l_\text{max}} \frac{1}{l!} \left \{ 
   \left( F^{(l)}(\bar{x})- \gamma^{(l)}(\bar{x})\bar{p}\right ) 
    (\Delta_{n+l,m-1}-\Delta_{l,0}\Delta_{n,m-1})
   - \gamma^{(l)}(\bar{x}) (\Delta_{n+l,m}-\Delta_{l,1}\Delta_{n,m-1})
  \right \},
\label{eq:FokkerPlanck}\end{align} 
which corresponds to the classical part ($\hbar \rightarrow 0$) of the set of dynamical
equations (\ref{eq:dglX})-(\ref{eq:dD03dt}).

\section{Derivation of dynamical equations from a quantum Lindblad master equation}\label{sec:basics}
To describe the cantilever as a quantum system,
we model it as a 1D quantum harmonic oscillator of mass $M$ that moves in a potential
$V(\hat{x})$ and is subject to noise. The density matrix $\rho$ obeys the
Master equation in Lindblad form \cite{linblad:cmp76},
\begin{align} 
  \dot{\rho} &= -\frac{ i}{\hbar} [\hat{H},\rho] - \sum_k 
  \left (
   \hat{J}_k^\dagger  \hat{J}_k \rho + \rho  \hat{J}_k^\dagger
               \hat{J}_k
  -2  \hat{J}_k \rho  \hat{J}_k^\dagger 
  \right )
\\
  \hat{H} &= \frac{ \hat{p}^2}{2M} + V(\hat{x}),
\end{align} 
where $ \hat{J}_k$ are jump operators. Force 
\begin{align} 
   F(\hat{x}) =-\nabla V \quad &=
   -k \hat{x} + F_\text{dr}(t) + F_\text{sf}(\hat{x})
\label{eq:Fcons}\end{align} 
contains all conservative
forces that act on the cantilever. This includes the elastic force $-k x$ of the cantilever
itself, the homogeneous two-frequency driving force $F_\text{dr}(t)$,
and the conservative part $F_\text{sf}$ of the force
exerted by the sample surface.
The time evolution of the expectation value
of an operator $\hat{A}(t)$, for which we admit an explicit time dependence, is given by
\begin{align} 
  \frac{ d}{dt}\langle A \rangle &= 
   \langle \partial_t \hat{A} \rangle 
   -\frac{ i}{\hbar} \langle  [\hat{A},\hat{H}] \rangle 
   - \sum_k
    \left \langle 
  [ \hat{A}, \hat{J}_k^\dagger]  \hat{J}_k 
  + \hat{J}_k^\dagger [ \hat{J}_k , \hat{A}] \right  \rangle .
\label{eq:masterEq}\end{align} 

Our goal is to derive the equations of motion for mean
Position $\bar{x}(t) = \langle \hat{x} \rangle $, mean momentum 
$\bar{p}(t) = \langle \hat{p} \rangle $, as well as for their
variances $\Delta x$ and $\Delta p$. 
We define a family of symmetric
correlation functions  $\Delta_{n,m}=\langle \hat{\Delta}_{n,m} \rangle $, with
\begin{align} 
  \hat{\Delta}_{n,m} &= \frac{ 1}{2} 
     (\delta\hat{x}^n \delta\hat{p}^m +
                 \delta\hat{p}^m \delta\hat{x}^n )
\label{def:Deltanm}\\
  \delta\hat{x} &= \hat{x}-\bar{x}
\label{eq:defDeltaX}\\
    \delta\hat{p} &= \hat{p}-\bar{p} .
\label{eq:defDeltaP}\end{align} 
There are two prominent
noise sources: (i) the damping force due to the finite quality factor
$Q$ of the cantilever, and (ii) the dissipative force associated with
the surface interaction.
Damping of the
cantilever can be modeled by adding a force $F_Q=-\gamma_Q p$
with damping constant $\gamma_Q=\frac{ \omega_0}{Q}$ \cite{Platz2013}.
In open quantum systems, such a friction force is described through a
Caldeira-Leggett model \cite{Caldeira1983}
in Lindblad form, see Eq.~(5.117) of
Ref.~\cite{HornbergerDecoherence}, for instance. In this model, 
the jump operator takes the form
\begin{align} 
  \hat{J}_Q &= \frac{ \sqrt{\gamma_Q}}{2} \left (   \frac{ p_\text{th}}{\hbar} \hat{x} +
   \frac{ i}{p_\text{th}} \hat{p} \right ),
\end{align} 
and the Hamiltonian is
modified by adding a term
\begin{align} 
  \hat{H}_Q &= 
  \frac{ \gamma_Q}{4} (\hat{x}\hat{p}+\hat{p}\hat{x}).
\end{align} 
Here, $p_\text{th}=2\sqrt{M k_B T}$ corresponds to the momentum
uncertainty in thermal equilibrium. Putting this together, the full
action of the Caldeira-Leggett model can be written in form of a super-operator,
\begin{align} 
  {\cal L}_Q\hat{A} &= -\frac{i}{\hbar}[\hat{A},\hat{H}_Q]
   -   [ \hat{A}, \hat{J}_Q^\dagger]  \hat{J}_Q 
   - \hat{J}_Q^\dagger [ \hat{J}_Q , \hat{A}] 
\\ &=
  -\frac{ \gamma_Q}{4p_\text{th}^2}
   \left [\hat{p},  [\hat{p},\hat{A}  ] \right ]
  -\frac{ \gamma_Q p_\text{th}^2}{4\hbar^2}
  \left [\hat{x},  [\hat{x},\hat{A} ] \right ]
  - \frac{ i\gamma_Q}{2\hbar} \left (
  [\hat{A},\hat{x}]\hat{p} +\hat{p} [\hat{A},\hat{x}]
  \right ).
\end{align} 

To describe the dissipative part of the surface-tip interaction, we
adopt the model of Ref.~\cite{Platz2013}, where the dissipative part
takes the form of a drag force that depends on the distance from the
sample, $F_{\text{dis}} =- p  \gamma(x)$. For most analytical calculations, we
will
keep the position-dependent dissipation rate $ \gamma(x)$ general.
However, for numerical evaluations, we will follow
Ref.~\cite{Platz2013} and use an exponential-decay model,
\begin{align} 
   \gamma(x) &= \gamma_0\exp \left(
  -\frac{ x+h}{x_\gamma} \right ),
\label{eq:PlatzDissip}\end{align}
where $\gamma_0$ is the dissipation rate at the sample surface and
$x_\gamma$ the length scale on which the dissipative force drops off. 
In a quantum treatment,
$\dot{x}$ has to be replaced by $\hat{p}/m$. One also has to write the
operator product in a symmetric way to ensure that the force operator
is hermitian. We therefore seek to generate a dissipative force operator
of the form
\begin{align} 
    \hat{F}_{\text{dis}} &=- \frac{1}{2}\left (
   \hat{p} \, \gamma(\hat{x})   + \gamma (\hat{x})\hat{p}
   \right ).
\label{eq:Fdis}\end{align} 
We remark that this reduces to the previous case for 
$\gamma(x)=\gamma_Q $.
We found that it is possible to describe this process through a
Lindblad master equation with
\begin{align} 
  \hat{J}_\text{dis} &=
  \frac{p_\text{th}\sqrt{\tau_0}}{4 \hbar} \Gamma(\hat{x})+
   \frac{ i}{p_\text{th}\sqrt{\tau_0}} \hat{p} 
\label{eq:Jdis}\\
  \hat{H}_\text{dis} &= 
  \frac{ 1 }{4} 
  \left ( \hat{p}
  \Gamma (\hat{x}) +  \Gamma (\hat{x})\hat{p}
  \right ),
\end{align} 
where $\Gamma(x)$ is the anti-derivative of $\gamma(x)$ and $\tau_0$
a time scale that will be discussed below. Note that $\tau_0$ 
may be a function of time. We can then again describe the full action
of the dissipative force in terms of a super-operator,
\begin{align} 
  {\cal L}_\text{dis}\hat{A} &= -\frac{i}{\hbar}[\hat{A},\hat{H}_\text{dis}]
   -   [ \hat{A}, \hat{J}_\text{dis}^\dagger]  \hat{J}_\text{dis} 
   - \hat{J}_\text{dis}^\dagger [ \hat{J}_\text{dis} , \hat{A}] 
\\ &=
  -\frac{ 1}{\tau_0 p_\text{th}^2}
   \left [\hat{p},  [\hat{p},\hat{A}  ] \right ]
  -\frac{ \tau_0 p_\text{th}^2}{16\hbar^2}
  \left [\hat{\Gamma},  [\hat{\Gamma},\hat{A} ] \right ]
  - \frac{ i}{2\hbar} \left (
  [\hat{A},\hat{\Gamma}]\hat{p} +\hat{p} [\hat{A},\hat{\Gamma}]
  \right ).
\end{align} 

We now have to evaluate the equations of motion,
\begin{align} 
  \frac{ d}{dt}\langle A \rangle &= 
   \langle \partial_t \hat{A} \rangle 
   -\frac{ i}{\hbar} \langle  [\hat{A},\hat{H}] \rangle 
  +  \left \langle 
    {\cal L}_Q\hat{A} +{\cal L}_\text{dis}\hat{A}
     \right  \rangle ,
\label{eq:masterEq2}\end{align} 
for the correlation functions (\ref{def:Deltanm}), which include an
explicit time dependence in their definition through $\bar{x}(t)$
and $\bar{p}(t)$.
We start with mean position and momentum, which obey 
\begin{align} 
  \dot{\bar{x}} &= \frac{ \bar{p}}{M}
\\
    \dot{\bar{p}} &= \langle \hat{F}_\text{tot} \rangle -\gamma_Q \bar{p},
\end{align}
where a dot denotes the time derivative $d/dt$, and
$\hat{F}_\text{tot} = F(\hat{x})+ \hat{F}_\text{dis}$. 
For the correlation functions, we find
\begin{align} 
  \dot{\Delta}_{n,m} &=
   \frac{ n}{M} 
  \Delta_{n-1,m+1}
  -\frac{i\hbar}{4M}n(n-1) 
  \left \langle
   [ \delta\hat{x}^{n-2}, \delta\hat{p}^{m}] \right \rangle 
  + \frac{i}{2\hbar} \big \langle 
   \delta\hat{x}^{n} [V, \delta\hat{p}^{m}] 
 +   [V, \delta\hat{p}^{m}] \delta \hat{x}^{n} \big \rangle 
  -m \langle \hat{F}_\text{tot} \rangle  \Delta_{n,m-1}
\nonumber \\ &\hspace{4mm}
 -m\gamma_Q \Delta_{n,m}
    + \frac{\gamma_Q p_\text{th}^2}{4} m(m-1)\Delta_{n,m-2}
  +\frac{\hbar^2}{4 p_\text{th}^2} 
  \left (\gamma_Q +\frac{4}{\tau_0}\right) n(n-1) \Delta_{n-2,m}
\nonumber \\ &\hspace{4mm}
  +\frac{ i\hbar\gamma_Q}{4}nm 
   \left\langle [\delta\hat{x}^{n-1},\delta\hat{p}^{m-1}] 
  \right \rangle 
  -\frac{ \tau_0 p_\text{th}^2}{16\hbar^2}
  \left \langle 
  \left [\hat{\Gamma},  [\hat{\Gamma}, \hat{\Delta}_{n,m}]  \right ]
  \right \rangle 
  + \frac{ i}{2\hbar} \left \langle 
  [\hat{\Gamma},\hat{\Delta}_{n,m}]\hat{p} +\hat{p} [\hat{\Gamma},\hat{\Delta}_{n,m}]
  \right \rangle .
\end{align} 
This result is exact, but its usefulness hinges on our ability to
evaluate commutators of powers of $ \hat{p}$ and functions of
$ \hat{x}$. This is accomplished in appendix
\ref{sec:commutator}. Result~(\ref{eq:VpnCommutator})
states that a commutator of the form
$ [V(\hat{x}), \hat{p}^n ]$ can be written in terms of Euler polynomials
with an argument that contains $\hat{p} $ and the derivative
operator $\partial/\partial \hat{x}$, which acts on $V(\hat{x})$.
In particular, if the function $V(\hat{x})$ is a polynomial, the
commutator corresponds to 
\begin{align} 
  \langle [\delta \hat{x}^m, \delta\hat{p}^n] \rangle &=
  i \sum_{l=1}^{\text{min}(m,n)} 
   C_{n,m,l} \hbar^l
  \Delta_{m-l,n-l},
\label{eq:xPowerpPowerCommSimplified}\end{align} 
where coefficients $C_{n,m,l}$ are defined through
Eq.~(\ref{eq:xPowerpPowerComm}). Hence, it can be expressed in terms of 
correlation functions (\ref{def:Deltanm}). 
However, potential $V(\hat{x})$ and factor $\Gamma(\hat{x})$ will
generally not be of a polynomial form, so that the equations of motion will not
form a closed set of equations. It is therefore necessary to make
approximations. 

With localization approximation (\ref{eq:locApprox}), the expectation value of all quantities
appearing in the equations of motion can be expressed in terms of
correlation functions (\ref{def:Deltanm}). For instance,
\begin{align} 
   \langle F(\hat{x}) \rangle &= 
    \sum_{n=1}^{n_\text{max}} \frac{ F^{(n)}(\bar{x})}{n!}
        \Delta_{n,0}
\\
   \langle F_\text{dis}(\hat{x}) \rangle &= - \frac{ 1}{2}
   \sum_{n=1}^{n_\text{max}} \frac{ \gamma^{(n)}(\bar{x})}{n!}
  \left \langle 
   \hat{p} (\hat{x}-\bar{x})^n+  (\hat{x}-\bar{x})^n\hat{p}
   \right \rangle 
\\ &= -\sum_{n=1}^{n_\text{max}} \frac{ \gamma^{(n)}(\bar{x})}{n!}
  \left (
    \Delta_{n,1} + \bar{p}\Delta_{n,0}
  \right ).
\end{align} 
We note that elastic force $-kx$,
cantilever damping $\gamma_Q$, and the
homogeneous driving force $F_\text{dr}(t)$ are taken into account
exactly. This is because their Taylor series terminates after the
first term, so that the above approximation leaves these forces unaffected.

Using similar expansions up to third order ($n_\text{max}=3$) for all terms in the equation of motions, and
treating terms $\Delta_{n,m}$ as of order $\epsilon^{n+m}$ in the
deviations from mean values, we arrive at the following equations,
\begin{align} 
 \frac{ d\bar{x}}{dt} &= \frac{ \bar{p}}{M}
\label{eq:dglX}\\
   \frac{d\bar{p}}{dt} &: \text{ see Eq.~(\ref{eq:pbarDot})}
\\
  \frac{ d\Delta_{2,0}}{dt} &=
   \frac{2}{M}\Delta_{1,1} + \frac{{\color{blue} \hbar}^2}{2p_\text{th}^2}
   \left (\gamma_Q + \frac{ 4}{\tau_0 }\right )
\label{eq:delxEqn}\\
  \frac{ d \Delta_{1,1} }{dt} &=
    \frac{1}{M}\Delta_{0,2}
     -(\gamma_Q +\gamma(\bar{x}) )\Delta_{1,1}
  + \Delta_{2,0}\left (  F'(\bar{x}) 
  -  \bar{p}\, \gamma '(\bar{x}) \right )
  -  \gamma '(\bar{x}) \Delta_{2,1} +
   \frac{ 1}{2}\Delta_{3,0} \left (  F''(\bar{x}) 
  -  \bar{p}\, \gamma ''(\bar{x}) \right )
\\
    \frac{d \Delta_{0,2}  }{dt}&=
  2\left ( F'(\bar{x}) -\bar{p} \gamma'(\bar{x}) 
  \right ) \Delta_{1,1}
  -2\gamma_Q \left (
  \Delta_{0,2} - \frac{ p_\text{th}^2}{4}
  \right )
  -  2  \gamma (\bar{x}) \Delta_{0,2}
  - \frac{ {\color{blue} \hbar}^2}{2} \gamma''(\bar{x}) 
  + \left ( F''(\bar{x}) -\bar{p} \gamma''(\bar{x}) 
  \right ) \Delta_{2,1}
\nonumber \\ &\hspace{4mm} 
  - 2 \gamma'(\bar{x}) \Delta_{1,2}
   +  \frac{ p_\text{th}^2 \tau_0}{8}
   \left (\gamma^2(\bar{x})+  
  \Delta_{2,0}  ( \gamma '(\bar{x})^2 + \gamma (\bar{x}) \gamma
               ''(\bar{x})  )
  + \left (
    \gamma '(\bar{x}) \gamma ''(\bar{x})
  + \frac{ 1}{3} \gamma(\bar{x}) \gamma '''(\bar{x})
  \right ) \Delta_{3,0}
    \right )
\label{dgl:dp2}\\
  \frac{ d\Delta_{3,0}  }{dt}&= \frac{3}{M} \Delta_{2,1}
\\
  \frac{d  \Delta_{2,1} }{dt} &= 
   \frac{2 \Delta _{1,2}}{M}
 -(\gamma_Q+ \gamma(\bar{x}))\Delta_{2,1} 
 +\Delta _{3,0} \left(F'(\bar{x})
  -\bar{p} \gamma '(\bar{x})\right)
\\
 \frac{d  \Delta_{1,2} }{dt} &= 
  \frac{\Delta _{0,3}}{M}
  -2 (\gamma _Q+\gamma (\bar{x}))\Delta _{1,2} 
   + 2 \Delta _{2,1} 
  \left(F'(\bar{x})-\bar{p} \gamma '(\bar{x})\right)
  + \Delta _{2,0} 
  \left(\frac{ p_\text{th}^2\tau_0}{4}\gamma' (\bar{x}) 
   \gamma(\bar{x}) - {\color{blue} \hbar}^2 \gamma'''(\bar{x}) 
   \right)
  - {\color{blue} \hbar}^2 \gamma '(\bar{x})
\nonumber \\ &\hspace{4mm} 
  + \Delta _{3,0} 
   \frac{ p_\text{th}^2\tau_0}{8}
   \left(  \gamma(\bar{x})  \gamma''(\bar{x})
  +  \gamma'^2(\bar{x})\right)
\\
  \frac{ d\Delta_{0,3}}{dt} &= 
     -3 \Delta _{0,3} (\gamma_Q+\gamma(\bar{x}))
  +3 \Delta _{1,2}
   \left(F'(\bar{x})-\bar{p} \gamma'(\bar{x})\right)
  +\frac{1}{2}  {\color{blue} \hbar}^2
   \left(F''(\bar{x})-\bar{p} \gamma''(\bar{x})\right)
-2  {\color{blue} \hbar}^2 \Delta _{1,1} \gamma'''(\bar{x})
\nonumber \\ &\hspace{4mm} 
  +\frac{3}{8} \tau _0 p_{\text{th}}^2
   \left (
  2 \Delta_{1,1} \gamma (\bar{x}) \gamma'(\bar{x})
  +\Delta _{2,1} \left(\gamma '(\bar{x})^2
  +\gamma(\bar{x}) \gamma''(\bar{x})\right)
  \right ).
\label{eq:dD03dt}\end{align} 
These equations represent the main theoretical result of this work. We
have verified that, except for two types of terms, they agree with the corresponding equations for a
classical model based on the Fokker-Planck equation, which is derived
in appendix \ref{sec:FokkerPlanck}. The two types of terms in which
the two models differ are (i) terms that depend on $\hbar$, and (ii)
terms that depend on derivatives of $\gamma(x)$. We will now discuss these differences,

For better identification terms of type (i), we have printed all occurrences of $\hbar$
in blue. These terms correspond to genuine quantum dynamics.
The first occurence in Eq.~(\ref{eq:delxEqn}) corresponds to
the position uncertainty relation $\Delta x \sim \hbar/\Delta p$ if
the momentum uncertainty is equal to the thermal momentum
$p_\text{th}$.
The only other occurence in the equations for the second-order
variances depends on the curvature $\gamma''(\bar{x})$ of the
dissipative part of the surface force. It is interesting to observe
that applying a position-dependent dissipative force may an effective
way to observe differences between classical and quantum dynamics of
a localized system. By
comparison, quantum dynamics induced by the conservative part of the
force only 
appear in the
dynamical equation for the momentum skewness $\Delta_{0,3}$. 

We now turn to terms of type (ii) and the
role of time scale $\tau_0$. We start by considering the
stationary solution for the case that $F=0=\gamma_Q$. The equations of
motion then have a quasi-stationary solution of the form
$\bar{p}=0$, $\bar{x}$ constant, as well as
\begin{align} 
  \Delta p &=  \frac{ p_\text{th} }{4} \sqrt{\tau_0\gamma(\bar{x})}
\\
  \Delta_{1,1} &= \frac{ p_\text{th}^2 \tau_0 }{16m} 
\\
  \Delta x &= \sqrt{
   t \left (  \frac{ p_\text{th}^2 \tau_0 }{8M^2} 
    + \frac{ 2\hbar^2}{p_\text{th}^2 \tau_0}
   \right ) 
   },
\end{align} 
and $\Delta_{n,m}=0$ for the third-order correlation functions.
If we want to ensure that the momentum uncertainty $\Delta p$ is equal to
$ p_\text{th}/2$ in this case, we have to set $\tau_0=4 /\gamma(\bar{x})$.

A second reason why $\tau_0$ should be equal to $4 /\gamma(\bar{x})$
is the comparison with the classical Fokker-Planck equation
(\ref{eq:FokkerPlanck}). In the classical limit ($\hbar=0$) and for
constant $\bar{x}$, the two sets of equations only agree if $\tau_0$ is chosen
in this way. 

However, even for $\tau_0=4 /\gamma(\bar{x})$ and in the classical
limit, the Fokker-Planck equation differs from the results above if
$\gamma(\bar{x}(t))$ varies with $\bar{x}(t)$, and these are the terms
of type (ii). In the quantum derivation, these terms are a direct
consequence of model (\ref{eq:Jdis}) for inhomogeneous quantum
dissipative forces. This model appears to be the only one where the
jump operator $\hat{J}_\text{dis}$ is linear in the momentum
operator. Hence, as long as the dissipative force is created by
systems that are Markovian (this is the underlying assumption of the
Lindblad form), the derivative terms are needed for
consistency. Furthermore, as discussed in appendix
\ref{sec:FokkerPlanck}, the absence of derivatives of $\gamma(x)$ in
the classical equations (\ref{eq:FokkerPlanck}) is merely a
consequence of an approximation. A refined Fokker-Planck model would
likely generate similar terms, so that we believe that the presence 
of type (ii) terms is physically well justified.

\section{Reconstructing a classical probability distribution from
  correlation functions}\label{sec:reconstruction}
We consider an $N$-dimensional phase space $\mathds{R}^N$ with $\boldsymbol{r}$ denoting an
element of this space. In our case, $\boldsymbol{r} = (x,p)$ is the 2D phase
space of a particle with one degree of freedom. A probability
distribution $\rho(\boldsymbol{ r})$ is a real positive function that is
normalized to unity, 
\begin{align} 
  1 &= \int d^N r\, \rho(\boldsymbol{r}).
\end{align} 
The $n$th moment around a point $\bar{\boldsymbol{r}} \in \mathds{R}^N$ is defined as
\begin{align} 
  \mu_{n} &= \int d^N r\, (\boldsymbol{ r}-\bar{\boldsymbol{r}})^n \rho(\boldsymbol{r}).
\end{align} 
Here and in the following, we will employ a multi-index notation: for
a tuple $n=(n_1, \cdots , n_N)$ of $k$ integer numbers, we set
\begin{align} 
   |n| &= n_1+n_2+\cdots +n_N
\\
  n! &= n_1! \, n_2!  \cdots n_N!
\\
  \boldsymbol{r}^{\, n} &= r_1^{n_1} r_2^{n_2}\cdots r_N^{n_k}
\\
  D^n f(\boldsymbol{ r}) &=
                    \frac{ \partial^{|n|}f}{\partial_1^{n_1}\cdots \partial_N^{n_N}}
\end{align} 
In this appendix, we answer the following question: Given the moments $\mu_n$ of a
probability distribution $\rho(\boldsymbol{ r})$, can we reconstruct
$\rho$?

To describe asymmetric probability distributions, Azzalini and Valle 
\cite{Azzalini1996} introduced a multi-variable
extension of the skew-normal distribution of the form
\begin{align} 
   \rho(\boldsymbol{r})&= \rho_0(\boldsymbol{r}) \Phi(\boldsymbol{r}),
\end{align}   
where $\rho_0$
corresponds to a Gaussian distribution (\ref{eq:GaussianMultiVar}),
with $C$ a positive symmetric $N\times N$ matrix.

In Ref.~\cite{Azzalini1996}, $\Phi$ is a specific
function, but we consider $\Phi$ as an unknown function that needs to be reconstructed using
the moments. We denote the moments of the Gaussian probability distribution
by  $\mu_n^{(0)}$ and assume that $\Phi$ possesses a Taylor
expansion around $\bar{\boldsymbol{r}}$, which in multi-index notation takes the form
\begin{align} 
  \Phi(\boldsymbol{ r}) &= \sum_k \frac{ 1}{k!} D^k\Phi(\bar{\boldsymbol{r}}) \,
                   (\boldsymbol{r}-\bar{\boldsymbol{r}})^{\, k}.
\end{align} 
We then can express the moments around point $\bar{\boldsymbol{r}}$ as
\begin{align} 
  \mu_n &= \int d^Nr\, (\boldsymbol{ r}-\bar{\boldsymbol{r}})^{\,n} \, \rho_0(\boldsymbol{ r}) 
    \sum_k \frac{ 1}{k!} D^k\Phi(\bar{\boldsymbol{r}}) \, (\boldsymbol{r}-\bar{\boldsymbol{r}})^{\, k}
\\ &=
  \sum_k \frac{ 1}{k!} D^k\Phi(\bar{\boldsymbol{r}}) \mu_{n+k}^{(0)}.
\label{eq:reconstr}\end{align} 
This corresponds to a set of linear equations that enable us to
express, up to a given order, $D^k\Phi(\bar{\boldsymbol{r}})$ in terms of $\mu_n$.
Of particular interest is the case when $\bar{\boldsymbol{r}}$ corresponds to the
measured mean value. In this situation, moments $\mu_k$ directly
correspond to the correlation function $\Delta_{n,m}$ that we study in
the main part. 

For a Gaussian distribution, all odd moments vanish,
$\mu^{(0)}_{2n+1}=0$. The even moments are given by
\begin{align} 
  \mu^{(0)}_{i_1,i_2,\cdots,i_{2n}} &=
   \int d^Nr\, \rho_0(\boldsymbol{r}) (\boldsymbol{ r}-\bar{\boldsymbol{r}})_{i_1}
   \cdots  (\boldsymbol{ r}-\bar{\boldsymbol{r}})_{i_{2n}}
\\ &=
   \frac{ 1}{(2\pi)^{N/2}|C|^{\frac{ 1}{2}}}
   \int d^Nx\, 
   x_{i_1} \cdots  x_{i_{2n}}
   e^{-\frac{ 1}{2} 
   \boldsymbol{x}^T\cdot C^{-1} \cdot \boldsymbol{x}}
\\ &=
 \frac{1}{2^n n!} \sum_p
  C_{i_{p(1})i_{p(2)}} \cdots
   C_{i_{p(2n-1)} i_{p(2n)}}, 
\end{align} 
where the sum runs over all permutations $p$ of the indices 
$i_1, \cdots i_{2n}$. Explicitly, 
moments 2, 4, and 6 are given by
\begin{align} 
  \mu^{(0)}_{i_1,i_2} &= C _{i_1,i_2}
\\
   \mu^{(0)}_{i_1,i_2,i_3,i_4} &=
   C  _{i_1,i_4} C  _{i_2,i_3}+
  C_{i_1,i_3} C  _{i_2,i_4}
  +C _{i_1,i_2} C  _{i_3,i_4}
 \\
   \mu^{(0)}_{i_1,i_2,i_3,i_4,i_5,i_6} &=
   C _{i_1,i_6} C _{i_2,i_5} C_{i_3,i_4}
  +C _{i_1,i_5} C_{i_2,i_6} C _{i_3,i_4}
   +C_{i_1,i_2} C _{i_5,i_6} C_{i_3,i_4}
   +C _{i_1,i_6} C_{i_2,i_4} C _{i_3,i_5}
  +C _{i_1,i_4} C _{i_2,i_6} C_{i_3,i_5}
\nonumber \\ &\hspace{4mm}
   +C _{i_1,i_5} C_{i_2,i_4} C _{i_3,i_6}
  +C_{i_1,i_4} C _{i_2,i_5} C_{i_3,i_6}
   +C _{i_1,i_6} C_{i_2,i_3} C _{i_4,i_5}
   +C_{i_1,i_3} C _{i_2,i_6} C_{i_4,i_5}
  +C _{i_1,i_2} C_{i_3,i_6} C _{i_4,i_5}
\nonumber \\ &\hspace{4mm}
  +C_{i_1,i_5} C _{i_2,i_3} C_{i_4,i_6}
   +C _{i_1,i_3} C_{i_2,i_5} C _{i_4,i_6}
   +C_{i_1,i_2} C _{i_3,i_5} C_{i_4,i_6}
  +C _{i_1,i_4} C_{i_2,i_3} C _{i_5,i_6}
  +C_{i_1,i_3} C _{i_2,i_4} C_{i_5,i_6}.
\label{eq:mu06}\end{align} 
We now use these results and 
Eq.~(\ref{eq:reconstr}) to find the Taylor coefficients
$D^k\Phi(\bar{\boldsymbol{r}})$ up to order $k=3$. We start by writing down
explicitly the first four moments. $\mu_0=1$ expresses normalization
of probability,
\begin{align} 
   1 &= \mu_0 
\\ &= \Phi(\bar{\boldsymbol{r}}) \mu^{(0)}_0  + \frac{ 1}{(2)!}D^2\Phi(\bar{\boldsymbol{r}})
     \mu^{(0)}_2
\label{eq:moment0}\\ &=\Phi(\bar{\boldsymbol{r}}) + \sum_{i_1,i_2}
   \frac{ 1}{(i_1 i_2)!} (\partial_{i_1}\partial_{i_2}\Phi(\bar{\boldsymbol{r}})) C_{i_1i_2}.
\end{align} 
Here, notation $(i_1 i_2)!$ is a multi-index notation that is equal to
2! if both indices are equal, and 1 otherwise. Below, we will also use
$(i_1 i_2 i_3)!$ equal to 3! for all three indices equal, 2! if only
two are equal, and 1 otherwise.

Since $\bar{\boldsymbol{r}}$ is equal
to the mean position, $\mu_1=0$ by definition. Hence, the second
equation becomes
\begin{align} 
  0 &= D^1\Phi\, \mu^{(0)}_2 + \frac{ 1}{(3)!}D^3\Phi \, \mu^{(0)}_4 ,
\end{align} 
or with explicit indices,
\begin{align} 
  0 &= \sum_{i_2}(\partial_{i_2}\Phi)\, \mu^{(0)}_{i_1 i_2} 
    +\sum_{i_2,i_3,i_4} \frac{ 1}{(i_2 i_3 i_4)!} 
  ( \partial_{i_2}\partial_{i_3}\partial_{i_4}\Phi )\, 
  \mu^{(0)}_{i_1  i_2 i_3 i_4}
\\ &=
    \sum_{i_2}(\partial_{i_2}\Phi)\, C_{i_1 i_2} 
    +\sum_{i_2,i_3,i_4} \frac{ 1}{(i_2 i_3 i_4)!} 
  ( \partial_{i_2}\partial_{i_3}\partial_{i_4}\Phi )\, 
  (C  _{i_1 i_4} C  _{i_2 i_3}+
  C_{i_1 i_3} C  _{i_2 i_4}
  +C _{i_1 i_2} C  _{i_3 i_4}).
\end{align} 
Since the multi-factorial and the derivatives of $\Phi$ are totally
symmetric under exchange of indices $i_2,i_3,i_4$, we can reduce the
second factor to a sum over a single term,
\begin{align} 
  0 &=
    \sum_{i_2}(\partial_{i_2}\Phi)\, C_{i_1 i_2} 
    + 3 \sum_{i_2,i_3,i_4} \frac{ 1}{(i_2 i_3 i_4)!} 
  ( \partial_{i_2}\partial_{i_3}\partial_{i_4}\Phi )\, 
  C  _{i_1 i_4} C  _{i_2 i_3} .
\end{align} 
Multiplying this with $(C^{-1})_{i i_1}$ and summing over $i_1$ yields
\begin{align} 
  \partial_{i}\Phi(\bar{\boldsymbol{r}}) &= -3
   \sum_{i_2,i_3} \frac{ 1}{(i \, i_2 i_3 )!} 
  ( \partial_{i}\partial_{i_2}\partial_{i_3}\Phi )\, C  _{i_2 i_3} .
\label{eq:D1phiIntermediate}\end{align} 

The second-order moment equals the correlation matrix, 
$\mu_{i_1 i_2} = \langle (\boldsymbol{r}-\bar{\boldsymbol{r}})_{i_1}
(\boldsymbol{r}-\bar{\boldsymbol{r}})_{i_2} \rangle $. The equation for this moment reads
\begin{align} 
  \mu_2 &= \Phi(\bar{\boldsymbol{r}}) \mu^{(0)}_2  + \frac{ 1}{(2)!}D^2\Phi(\bar{\boldsymbol{r}})
     \mu^{(0)}_4 .
\label{eq:moment2}\end{align} 
Together with Eq.~(\ref{eq:moment0}), this provides us with a linear
set of equations for the $N^2+1$ coefficients $ \Phi(\bar{\boldsymbol{r}})$ and
$D^2\Phi(\bar{\boldsymbol{r}})$. This is easy to solve if we pick matrix $C$ in the
Gaussian distribution (\ref{eq:GaussianMultiVar}) such that
$\mu^{(0)}_2=\mu_2$. This is accomplished for the choice
$C_{i_1 i_2} = \mu_{i_1 i_2}$. 
Eqs.~(\ref{eq:moment0}) and (\ref{eq:moment2}) are then easily solved
by $ \Phi(\bar{\boldsymbol{r}})=1$ and $D^2\Phi(\bar{\boldsymbol{r}}) = 0$.

The equation for the third-order moment is given by
\begin{align} 
   \mu_3 &= 
   D^1\Phi\, \mu^{(0)}_4 + \frac{ 1}{(3)!}D^3\Phi \, \mu^{(0)}_6,
\end{align} 
or with explicit indices,
\begin{align} 
  \mu_{i_1 i_2 i_3} &=
  \sum_{i_4} (\partial_{i_4}\Phi)
    \mu^{(0)}_{i_1 i_2 i_3 i_4}
   + \sum_{i_4, i_5, i_6} \frac{1}{(i_4 i_5 i_6)!}
    (\partial_{i_4}\partial_{i_5}\partial_{i_6}\Phi)
    \mu^{(0)}_{i_1 i_2 i_3 i_4 i_5 i_6}
\end{align} 
We can exploit Eq.~(\ref{eq:D1phiIntermediate}) to turn this equation
into one that only contains the third-order derivatives 
$(\partial_{i_4}\partial_{i_5}\partial_{i_6}\Phi)$. The triple sum
contains factors like Eq.~(\ref{eq:mu06}), which are very
lengthy. However, the high degree of symmetry of all factors involved
enables us to recude it to
\begin{align} 
  \mu_{i_1 i_2 i_3} &= 6
  \sum_{i_4, i_5, i_6} \frac{1}{(i_4 i_5 i_6)!}
    (\partial_{i_4}\partial_{i_5}\partial_{i_6}\Phi)
   C_{i_1 i_4}C_{i_2 i_5}C_{i_3 i_6} .
\end{align} 
This equation is easily solved and lets us determine all Taylor
coefficients up to order 3,
\begin{align} 
    \partial_{i}\partial_{j}\partial_{k}\Phi(\bar{\boldsymbol{r}})
   &= \frac{ (i\, j\, k)!}{6} 
   \sum_{i_1, i_2, i_3} \mu_{i_1 i_2 i_3} (C^{-1})_{i\, i_1}
    (C^{-1})_{j \, i_2}  (C^{-1})_{k\, i_3}
\\
   \partial_{i}\Phi(\bar{\boldsymbol{r}})
   &= -\frac{ 1}{2}  \sum_{i_1, i_2, i_3} 
  \mu_{i_1 i_2 i_3} (C^{-1})_{i\, i_1}
    (C^{-1})_{i_2 \, i_3}.
\end{align} 
The full expression for $\Phi(\boldsymbol{r})$ is then given by Eq.~(\ref{eq:PhiResult}).

Turning to the two-dimensional case that is the subject of this study,
we use correlation matrix (\ref{eq:corrMat}), with
\begin{align} 
     C^{-1} &= \sigma_1^{-2}\sigma_2^{-2} \left ( \begin{array}{cc}
          \Delta_{02}   &  -\Delta_{11}\\   -\Delta_{11}&  \Delta_{20}
           \end{array} \right ) .
\end{align} 
Matrix $C$ has eigenvectors and eigenvalues
\begin{align} 
   \boldsymbol{e}_{\underline{1}} &= \frac{
                                    1}{\sqrt{2W(W-\Delta_{02}+\Delta_{20})}}
         \left ( \begin{array}{c}
          \Delta_{20}-\Delta_{02} +W   \\  2\Delta_{11}
           \end{array} \right ) 
  , \quad C\cdot \boldsymbol{e}_{\underline{1}} =
   \sigma_1^2 \boldsymbol{e}_{\underline{1}}  =
  \frac{1}{2}(\Delta_{20}+\Delta_{02}+W) \boldsymbol{e}_{\underline{1}} 
\\   \boldsymbol{e}_{\underline{2}} &= \frac{
                                    1}{\sqrt{2W(W+\Delta_{02}-\Delta_{20})}}
         \left ( \begin{array}{c}
          \Delta_{20}-\Delta_{02} -W   \\  2\Delta_{11}
           \end{array} \right ) 
  , \quad C\cdot \boldsymbol{e}_{\underline{2}} =
   \sigma_2^2 \boldsymbol{e}_{\underline{2}}  =
  \frac{1}{2}(\Delta_{20}+\Delta_{02}-W)
  \boldsymbol{e}_{\underline{2}} 
\\
  W &= \sqrt{4\Delta_{11}^2 +(\Delta_{20}-\Delta_{02})^2}.
\end{align} 
Here, $\sigma_1, \sigma_2$ are the variances along the directions of
the eigenvectors of $C$. The third-order moment has components
$\mu_{111}=\Delta_{30}$, $\mu_{222}=\Delta_{03}$, and
$\mu_{112}=\mu_{121}=\mu_{211}=\Delta_{21}$, as well as
$\mu_{221}=\mu_{212}=\mu_{122}=\Delta_{12}$.
Introducing tensor components
\begin{align} 
  T^{(1)}_1 &= \Delta_{30}+\Delta_{12}
\\
   T^{(1)}_2 &= \Delta_{03}+\Delta_{21}
\\
  T^{(3)}_1 &= \frac{1}{3} \Delta_{30}-\Delta_{12}
\\
     T^{(3)}_2 &=-\frac{1}{3} \Delta_{03}+\Delta_{21},
\end{align} 
we can express function $\Phi$ as
\begin{align} 
   \Phi(\boldsymbol{r}) &= 1 + \boldsymbol{S}\cdot \boldsymbol{R} 
  + \frac{|\boldsymbol{R}|^2}{8} \boldsymbol{T}^{(1)}\cdot
                          \boldsymbol{R} 
  + \frac{1}{8} T^{(3)}_1 (R_1^3-3R_1 R_2^2)
  + \frac{1}{8} T^{(3)}_2 (3R_2 R_1^2-R_2^3).
\end{align} 
In polar coordinates,
$R_1=R\cos\phi$ and $R_2=R\sin\phi$, tensors 
$\boldsymbol{T}^{(1)}$ and $\boldsymbol{T}^{(3)}$ describe terms that
vary like $\cos\phi$ and $\sin\phi$, or  $\cos3\phi$ and $\sin3\phi$,
respectively. Hence, $\boldsymbol{T}^{(1)}$ describes the direction in
which the extended tail of $\rho$ points, while $\boldsymbol{T}^{(3)}$ 
describes deformations of a triangular shape.

\section{Numerical Simulations} \label{sec:numDetails}
To solve the dynamical equations
(\ref{eq:dglX})-(\ref{eq:dD03dt})
numerically, we
consider a cantilever with a resonance frequency 
$f_0=\omega_0/(2\pi) = 300$ kHz, a spring constant of
$M\omega_0^2 = 40.0$ N/m, and a quality factor of $Q=400$. 
The cantilever starts from its equilibrium
position and from thermal equilibrium at room temperature (300 K).

To model the surface force, we follow
Platz {\em et al.} \cite{Platz2013} and assume
an exponentially decreasing dissipative force 
$F_\text{dis}=-\gamma(x) p$, where $\gamma(x)$ given by 
Eq.~(\ref{eq:PlatzDissip}), with
$x_\gamma=1.5$nm and $2\pi\gamma_0/\omega_0 \approx 0.065$.
For the conservative surface force, we employ a modification of the
van der Waals-Derjaguin-Muller-Toropov (DMT) model. In its original form,
the DMT force is given by
\begin{align} 
  F_\text{DMT}(x) &= \left \{  \begin{array}{cc}
           -\frac{H R}{6(a_0+x+h)^2}   &    x>-h
      \\  -\frac{H R}{6a_0^2} + \frac{4}{3} E^* \sqrt{R}
                                (-h-x)^{\frac{3}{2}} 
        & x< -h
           \end{array} \right . ,
\label{eq:WDMTmodelOriginal}\end{align} 
where $H =3.28\times 10^{-17} \text{kg m}^2\text{s}^{-2}$ is the
Hamaker constant, which is a measure for the van der Waals interaction 
energy between tip and surface. $R=10$ nm denotes the tip radius, and 
$a_0 =2.7\, $nm represents the intermolecular distance. 
$E^*= 1.5\,$GPa is the effective stiffness of the tip-sample
system. The piecewise definition of this force makes it unsuitable for
our purposes, since derivatives of Eq.~(\ref{eq:WDMTmodelOriginal})
are not well-defined at $x=-h$. We therefore employ a modified model,
which is continuously differentiable,
\begin{align} 
  F_\text{sf}(x) &= 
           -\frac{1.15 H R}{12}\frac{ 
  \left  (1+\tanh\left (\frac{ x+h}{L_f} \right ) \right )
     }{L_f^2+ (a_0+x+h)^2}   
   +\frac{ 1}{2}   \left  (1-\tanh\left (\frac{ x+h}{L_f} \right ) \right )
   \left (  -\frac{1.15 H R}{6(L_f^2 +a_0^2)}
   +\frac{4}{3} E^* \sqrt{R}
                                (L_f^2+(h+x)^2)^{\frac{3}{4}} 
   \right ) .
\label{eq:WDMTmodelSmooth}\end{align} 
Here, $L_f=a_0/4$ controls the smoothness of the transition between
van der Waals and surface region. A plot of model force
(\ref{eq:WDMTmodelSmooth}) and the original DMT model
(\ref{eq:WDMTmodelOriginal}) is shown in Fig.~\ref{fig:forcePlots}.

We have performed a series of numerical simulations of the full
dynamical equations (\ref{eq:dglX})-(\ref{eq:dD03dt}) with the 
parameters for the surface-tip interaction as given above. 
In all
simulations, we have considered several special cases:
\begin{itemize}
\item
{\em Full equations}: The full set of equations
(\ref{eq:dglX})-(\ref{eq:dD03dt}) is simulated
\item 
{\em Variance limit}: Only second order correlation functions
$\Delta_{20},\Delta_{11},\Delta_{02}$ are taken into account;
third-order variances ($\Delta_{n,m} $ with $n+m=3$) are set to zero.
\item 
{\em Point-particle limit}: all correlation functions $\Delta_{nm}$
are assumed to vanish.
\item
{\em Reduced dissipative force}: To study the influence of the
dissipative force, we have run the simulations in a situation where
the dissipative surface force is reduced by a factor of $10^{-3}$.
\item
{\em No quantum terms}: All quantum terms (the blue terms in 
Eqns.~(\ref{eq:dglX})-(\ref{eq:dD03dt})) are set to zero.
\end{itemize}
In addition, we have performed numerical simulations that include
fourth-order correlation functions ($\Delta_{n,m} $ with $n+m=4$) to
verify that these terms can be ignored. These results
were affirmative and are not
presented in this paper.

In agreement with first-order perturbation theory
(see appendix \ref{sec:drivenPT}), we found two general
results in our simulations. 
(i) If one is only interested in studying mean position and momentum
  of the cantilever, the point-particle limit is appropriate. 
  Variances only have a small effect on their dynamics.
(ii) For standard AFM parameters, quantum terms can safely be
neglected. Since thermal variances are generally several orders of
magnitude larger than quantum uncertainties, our results do not
support claims in the literature
that AFM is quantum-limited.

The simulation supports the findings that we found in perturbation
theory: significant squeezing is only generated when the tip is in
contact with the sample.

\section{Perturbation theory of the driven cantilever}\label{sec:drivenPT}
If the tip-surface interaction is sufficiently small, the surface
forces can be treated as a perturbation. 
To derive a solution of Eqs.~(\ref{eq:dglX}) -
(\ref{eq:dD03dt}) to first order in perturbation theory,
we assume that the
unperturbed system is initially thermalized, i.e., mean position and
momentum follow a stationary trajectory, and the variances correspond
to a thermal equilibrium. The full unperturbed solution for driving
force (\ref{eq:twoFreqDrivingForce}) is then given by
 the unperturbed mean position $x_0(t)$ of
 Eq.~(\ref{eq:xUnperturbed}),
unperturbed mean momentum $p_0(t)=M \dot{x}_0$, as well as
\begin{align} 
  \Delta x_\text{th}^2 &:=\Delta_{20}^{\text{(no srfc)}} = \frac{ k_BT}{M\omega_0^2} 
  \left ( 1 + \left ( 1+ \frac{ 1}{Q^2} \right ) 
    \left ( \frac{\hbar\omega_0}{4 k_B T }
   \right )^2  \right )
\\
  \Delta_{02}^{\text{(no srfc)}} &= M k_BT   \left ( 1 + 
    \left ( \frac{\hbar\omega_0}{4 k_B T }
   \right )^2  \right ),
\end{align} 
and $\Delta_{n,m}=0$ else. For a typical cantilever, the ratio of ground state energy $\hbar
\omega_0$ and thermal energy $k_B T$ is in the order of
$10^{-8}$. Here and in the following, we will therefore neglect terms
of order $\hbar^2$ and only keep lowest-order quantum
contributions. We will also neglect terms of order $Q^{-1}$ since the
quality factor is typically in the order of $10^2$.
With this approximation, we find the usual result for
the thermal uncertainty of a classical oscillator, 
$ \Delta x_\text{th} = \sqrt{k_BT/M\omega_0^2}$.

To include the effect of surface forces,
we consider the following
dimensionless 9-component vector of first-order corrections,
\begin{align} 
  \vec{V} &= \left ( 
   \frac{ \Delta_{20}^{(1)} }{L^2},
     \frac{ \Delta_{11}^{(1)} }{\hbar} ,
    \frac{ \Delta_{02}^{(1)}  L^2}{\hbar^2},
  \frac{ \Delta_{30}^{(1)} }{L^3},
    \frac{ \Delta_{21}^{(1)} }{L \hbar} ,
 \frac{ \Delta_{12}^{(1)}  L}{\hbar^2},
    \frac{ \Delta_{03}^{(1)}  L^3}{\hbar^3},
  \frac{ x^{(1)} }{L} , \frac{p^{(1)}  L}{\hbar}
  \right ),
\end{align} 
where a superscript $(1)$ indicates a first-order perturbation term
and $L=\sqrt{\hbar/(M\omega_0)}$ is the ground state width.
The perturbative dynamical equations for this vector can be written as
\begin{align} 
  \partial_t \vec{V} &= M\cdot \vec{V} + \vec{ J},
\label{eq:pt2}\end{align} 
with matrix
\begin{align} 
  M &=
  \left(
\begin{array}{ccccccccc}
 0 & 2 \omega _0 & 0 & 0 & 0 & 0 & 0 & 0 & 0 \\
 -\omega _0 & -\gamma _Q & \omega _0 & 0 & 0 & 0 & 0 & 0 & 0 \\
 0 & -2 \omega _0 & -2 \gamma _Q & 0 & 0 & 0 & 0 & 0 & 0 \\
 0 & 0 & 0 & 0 & 3 \omega _0 & 0 & 0 & 0 & 0 \\
 0 & 0 & 0 & -\omega _0 & -\gamma _Q & 2 \omega _0 & 0 & 0 & 0
   \\
 0 & 0 & 0 & 0 & -2 \omega _0 & -2 \gamma _Q & \omega _0 & 0 & 0
   \\
 0 & 0 & 0 & 0 & 0 & -3 \omega _0 & -3 \gamma _Q & 0 & 0 \\
 0 & 0 & 0 & 0 & 0 & 0 & 0 & 0 & \omega _0 \\
 0 & 0 & 0 & 0 & 0 & 0 & 0 & -\omega _0 & -\gamma _Q \\
\end{array}
\right),
\end{align} 
and inhomogeneity components $J_4=J_5=J_8=0$, and
\begin{align} 
J_1 &= -\frac{\hbar \gamma (\bar{x})}{8M\omega_0 \Delta x_{\text{th}}^2}
\\
  J_2 &= -\frac{\Delta x_{\text{th}}^2 \left(
   F_{\text{sf}}'(x_0) -p_0 \gamma'(x_0)
   \right ) }{\hbar }
  -\frac{\hbar  \left(F_{\text{sf}}'(x_0)
  -p_0 \gamma '(x_0)
  +M \gamma _Q \gamma (x_0)
  \right)}{16 M^2 \omega
   _0^2 \Delta x_{\text{th}}^2}
\\
 J_3 &= 
   -\frac{2 M \omega _0 \Delta x_{\text{th}}^4 
  \left(\gamma '(x_0)^2
   +\gamma (x_0) \gamma''(x_0)\right)
  }{\hbar  \gamma (x_0)}
  + \frac{\hbar  \left(\gamma_Q 
  \left(F_{\text{sf}}'(x_0)-p_0 \gamma'(x_0)\right)
   +M \omega _0^2 \gamma (x_0)\right)
  }{8 M^2 \omega_0^3 \Delta x_{\text{th}}^2}
  -\frac{\hbar \left(\gamma '(x_0)^2
    -3 \gamma(x_0) \gamma''(x_0)\right)
   }{8 M \omega _0 \gamma (x_0)}
\\
  J_6 &=
  -\frac{4 \Delta x_{\text{th}}^4 \gamma '(x_0)}{L^3}
  +L \Delta x_{\text{th}}^2 \gamma ^{(3)}(x_0)
   +\frac{3}{4} L \gamma'(x_0)
\\
  J_7 &= 
   \frac{3 L \gamma_Q \gamma '(x_0)}{4 \omega _0}
  -\frac{L \left(F_{\text{sf}}''(x_0)-p_0 \gamma ''(x_0)\right)
  }{2 M\omega _0}
\\
  J_9 &=
  -\frac{L }{\hbar }  \left (
   F_{\text{sf}}(x_0) -p_0 \gamma (x_0)
  + \Delta x_{\text{th}}^2\left(F_{\text{sf}}''(x_0)  - p_0 \gamma''(x_0)\right)
  \right ).
\end{align} 
It is worthwhile to note that matrix $M$ is block-diagonal and only
couples correlation functions $\Delta_{nm}$ of the same order
$n+m$. Hence, squeezing and other modifications of correlation
functions must be generated through the inhomogeneity $\vec{J}(t)$.

The solution of Eq.~(\ref{eq:pt2}) for $\vec{V}(0)=0$ is given by
\begin{align} 
  \vec{V}(t) &= \int_0^t dt'\, 
  e^{M(t-t')}\cdot \vec{J}(t').
\end{align} 
This is best evaluated by using the eigenvalues of $M$. This matrix is
not hermitian, but it is not singular. We can therefore express any
vector in the form
\begin{align} 
  \vec{J}(t') &= \sum_{\alpha=1}^9 \vec{e}_\alpha \tilde{J}_\alpha(t'),
\end{align} 
where $\vec{ e}_\alpha$ are the eigenvectors of $M$.
Specifically, the relationship between the original components
$J_i$ and the expansion coefficients $\tilde{J}_\alpha$ is given by
\begin{align} 
\\
  \tilde{J}_1 &= \frac{J_9 \left(2 \omega _0-i \gamma _Q\right)}{2
   \sqrt{2} \omega _0}-\frac{i J_8}{\sqrt{2}}  
\\ &=
  -\frac{L}{\sqrt{2} \hbar } 
  \left ( 1- i \frac{ \gamma_Q}{2\omega_0}
  \right )
  \left (   F_{\text{sf}}(x_0)-p_0 \gamma (x_0)
  + \Delta x_{\text{th}}^2 
    \left(F_{\text{sf}}''(x_0)-p_0 \gamma ''(x_0)\right)
  \right )
\label{eq:Jtilde1} \\ 
  \tilde{J}_3 &= \frac{J_2 \gamma _Q}{\sqrt{2} \omega_0}
   +\frac{J_1}{\sqrt{2}}+\frac{J_3}{\sqrt{2}}
\\ &=
  -\frac{\sqrt{2} M \omega _0 \Delta x_{\text{th}}^4
   \left(\gamma '(x_0)^2+\gamma (x_0)\gamma''(x_0)\right)
  }{\hbar  \gamma (x_0)}
-  \frac{ \gamma _Q 
  \left ( \Delta x_{\text{th}}^2
   -\frac{ L^4}{ 16\Delta x_{\text{th}}^2} \right )
   \left( F_{\text{sf}}'(x_0)-p_0 \gamma'(x_0)\right)
    }{\sqrt{2} \hbar \omega _0 }
  -\frac{\hbar 
   \left(\gamma '(x_0)^2-3 \gamma (x_0) \gamma'(x_0)\right)
   }{8 \sqrt{2} M \omega _0 \gamma (x_0)}
\\
  \tilde{J}_4 &=
\frac{\sqrt{3} J_3
   \left(\omega _0-i \gamma _Q\right)}{4 \omega
   _0}-\frac{\sqrt{3} J_2 \left(\gamma _Q+2 i \omega
   _0\right)}{4 \omega _0}-\frac{1}{4} \sqrt{3} J_1
\\
  &= \frac{\sqrt{3} \hbar  
   \left(4 M \omega _0^2 \gamma (x_0)
  +\left(3 \gamma _Q+2 i \omega _0\right) 
   \left(F_{\text{sf}}'(x_0)-p_0 \gamma'(x_0)\right)\right)
   }{64 M^2 \omega _0^3 \Delta x_{\text{th}}^2}
  +\frac{\sqrt{3}\Delta x_{\text{th}}^2 
  \left(\gamma_Q+2 i \omega _0\right)
   \left(F_{\text{sf}}'(x_0)-p_0 \gamma '(x_0)\right)
   }{4 \omega _0 \hbar }
\nonumber \\ &\hspace{4mm}
  +\frac{i \sqrt{3} M \Delta x_{\text{th}}^4 
  \left(\gamma _Q+i \omega _0\right) 
  \left(\gamma'(x_0)^2+\gamma (x_0) \gamma''(x_0)\right)
   }{2 \hbar  \gamma (x_0)}
  +\frac{i \sqrt{3} \hbar  \left(\gamma _Q+i \omega _0\right) 
   \left(\gamma '(x_0)^2-3\gamma (x_0) \gamma ''(x_0)\right)
   }{32 M \omega _0^2\gamma \left(x_0\right)}
\\ 
  \tilde{J}_6 &= \frac{\sqrt{5} J_7 
   \left(2 \omega _0-i \gamma _Q\right)}{8 \omega_0}
   +\frac{\sqrt{5} J_5 
    \left(2 \omega _0-3 i \gamma_Q\right)}{8 \omega _0}
   +\frac{\sqrt{5} J_6 \left(\gamma _Q-i\omega _0\right)}{4 \omega _0}-\frac{1}{4} i \sqrt{5}
   J_4
\\ &=
  -\frac{\sqrt{5}(\gamma_Q-i\omega_0) }{L^3\omega_0}\Delta x_{\text{th}}^4 \gamma '(x_0)
  +\frac{\sqrt{5}(\gamma_Q-i\omega_0)}{4\omega_0} L \Delta x_{\text{th}}^2 \gamma ^{(3)}(x_0)
\nonumber \\ &\hspace{4mm}
  + \frac{\sqrt{5} L  }{16 M \omega _0^2}
  \left(  (i\gamma_Q-2\omega_0)(F_{\text{sf}}''(x_0) - p_0 \gamma ''(x_0))
   +3 (2\gamma_q-i\omega_0)M \omega _0 \gamma'(x_0)\right)
\\
  \tilde{J}_8 &= \frac{J_7
   \left(2 \omega _0-3 i \gamma _Q\right)}{8 \omega _0}-\frac{3
   J_6 \left(\gamma _Q+i \omega _0\right)}{4 \omega _0}+\frac{3
   i J_5 \left(\gamma _Q+2 i \omega _0\right)}{8 \omega
   _0}+\frac{i J_4}{4}
\\ &=
  \frac{3 \Delta x_{\text{th}}^4 \left(\gamma _Q+i \omega _0\right) \gamma '(x_0)
  }{L^3 \omega _0}
  -\frac{3 L \Delta x_{\text{th}}^2\gamma ^{(3)}(x_0)
   \left(\gamma _Q+i \omega _0\right)}{4\omega _0} 
\nonumber \\ &\hspace{4mm}
  -\frac{L   }{16 M \omega _0^2}
  \left(\left(2 \omega _0-3 i \gamma _Q\right)
   \left(F_{\text{sf}}''(x_0)-p_0 \gamma ''(x_0)\right)
  +3 M\omega _0 \left(2 \gamma _Q+3 i \omega _0\right)\gamma '(x_0)\right),
\end{align} 
and $\tilde{J}_2 = \tilde{J}_1^*$,   $\tilde{J}_5 = \tilde{J}_4^*$, 
$\tilde{J}_7 = \tilde{J}_6^*$, and $\tilde{J}_9 = \tilde{J}_8^*$.
For $\gamma_Q\ll \omega_0$, the eigenvalues of $M$ are 
approximately given by
\begin{align} 
  \lambda_\alpha \in 
  \left \{ 
    -\frac{\gamma _Q}{2}-i \omega _0,
   -\frac{\gamma _Q}{2}+i\omega _0,
  -\gamma _Q,
   -\gamma _Q-2 i \omega_0,
   -\gamma _Q+2 i \omega _0,
  -\frac{3 \gamma _Q}{2}-i \omega_0,
   -\frac{3 \gamma _Q}{2}+i \omega _0,
   -\frac{3 \gamma _Q}{2}-3 i\omega _0,
    -\frac{3 \gamma _Q}{2}+3 i \omega _0
   \right \},
\end{align} 
which leads to 
\begin{align} 
  \vec{V}(t) &=  \sum_{\alpha=1}^9 \vec{e}_\alpha
  \int_0^t dt'\,  e^{\lambda_\alpha (t-t')}  \tilde{J}_\alpha(t').
\label{eq:pertSolution}\end{align} 

From this expression, we can draw several conclusions.\\[2mm]
(i) First-order perturbative effects on {\em Mean position and momentum} is
described through terms involving $\tilde{J}_1$ and $\tilde{J}_2$ in
solution (\ref{eq:pertSolution}).
These terms are not affected by quantum effects. They are affected by thermal fluctuations through 
terms proportional to $\Delta x_\text{th}$ in Eq.~(\ref{eq:Jtilde1}),
but a numerical estimate shows that this influence is small, roughly
in the order of $10^{-4}$. Therefore, in agreement with numerical
simulations, we conclude that the point-particle approximation is
appropriate if only the position of the tip is measured.
\\[2mm]
(ii) Terms involving $\tilde{J}_6$ to $\tilde{J}_9$ describe the
influence of the surface force and quantum effects on
{\em skewness}. Numerical simulations show that the overall size of skewness
remains small, so that we do not discuss the details of this case.
\\[2mm]
(iii) The third-order expansion presented above is sufficient
to describe squeezing and skewness for up to 300 cycles of the
cantilever. For longer times, fourth-order terms (coupling to
variances $\Delta_{nm}$ with $n+m=4$) can have a strong influence on
squeezing and skewness. We have analyzed the corresponding coupling
numerically, but since it is not relevant for normal AFM time scales,
we do not discuss it here. 
\\[2mm]
(iv) {\em Squeezing} of second-order variances
is the most interesting case, since it may be
observable and exhibits the largest contributions due to quantum dynamics.
Squeezing is introduced through terms involving $\tilde{J}_3$ to $\tilde{J}_5$.
To first order in $\gamma_Q$, the corresponding eigenvectors of $M$
are given by
\begin{align} 
  \vec{ e}_3 &= \left (\frac{1}{\sqrt{2}},-\frac{\gamma _Q}{2 \sqrt{2} \text{$\omega
   $0}},\frac{1}{\sqrt{2}},0,0,0,0,0,0\right)
\\
  \vec{ e}_4 &= \left (-\frac{1}{\sqrt{3}}-\frac{i \gamma _Q}{\sqrt{3} \omega _0},-\frac{\gamma _Q}{2 \sqrt{3}
   \omega _0}+\frac{i}{\sqrt{3}},\frac{1}{\sqrt{3}},0,0,0,0,0,0\right),
\end{align} 
and $\vec{e}_5 = \vec{ e}_4^*$. In this expression, the first three
components correspond to position variance, $\Delta_{11}$, and
momentum variance, respectively. For brevity, we will only discuss the
position variance, for which Eq.~(\ref{eq:pertSolution}) yields
\begin{align} 
   \Delta_{20}^{(1)} &= \int_0^t dt' e^{\gamma _Q (t'-t)} \Bigg [
  \frac{\gamma _Q 
  \left(\frac{ L^4}{16}-\Delta x_{\text{th}}^4\right) 
  }{2 M \omega _0^2 \Delta x_{\text{th}}^2}
  \left(F_\text{sf}'(x_0(t'))-p_0(t') \gamma'\left(x_0(t')\right)\right)
\nonumber \\ &\hspace{4mm}
   -
  \left(\Delta x_{\text{th}}^4-\frac{ 3 L^4}{16}\right) 
   \gamma''(x_0(t'))
   -\left(\frac{ L^4}{16}+ \Delta x_{\text{th}}^4\right) 
  \frac{ \gamma '(x_0(t'))^2  }{ \gamma \left(x_0(t')\right)}
  \Bigg ]
\nonumber \\ &\hspace{4mm}
  + \int_0^t dt' 
  \frac{ e^{\gamma _Q (t'-t)} \cos (2 \omega _0(t-t')) 
   }{\Delta x_{\text{th}}^2}
   \Bigg [
   \Delta x_{\text{th}}^2 
  \left(\Delta x_{\text{th}}^4-\frac{ 3 L^4}{16}\right) 
  \gamma ''(x_0(t'))
   - \frac{ L^4 }{8}\gamma(x_0(t'))
\nonumber \\ &\hspace{4mm}
-\frac{ \gamma_Q}{2M \omega _0^2} \left(\frac{ L^4}{16}- \Delta x_{\text{th}}^4\right)
   \left(F_\text{sf}'(x_0(t'))-p_0(t') \gamma'(x_0(t'))\right)
   +\Delta x_{\text{th}}^2
   \left(\frac{ L^4}{16}+ \Delta x_{\text{th}}^4\right) 
   \frac{ \gamma'(x_0(t'))^2}{\gamma(x_0(t')) }
  \Bigg ]
\nonumber \\ &\hspace{4mm}
  -\int_0^t dt' 
     \frac{ e^{\gamma _Q \left(t'-t\right)} 
  \sin (2\omega _0 (t-t')) }{ M \omega _0 \Delta x_{\text{th}}^2}
  \left [
  \left(\frac{ L^4}{16}+ \Delta x_{\text{th}}^4\right)
   \left(F_\text{sf}'(x_0(t'))-p_0(t') 
  \gamma'(x_0(t'))\right)
  +\frac{ L^4}{8} M \gamma _Q \gamma(x_0(t'))
  \right ]
\end{align} 
This expression shows that quantum effects are generally very
small. They enter through the ground state width $L$, which, at room
temperature, is about a factor of $10^{-4}$ smaller than the thermal
variance $\Delta x_{\text{th}}$ of the tip position.

To gain a better understanding of quantum effects, we have analyzed
this expression for the special case of a single-frequency driving
force ($F_2=0$ in Eq.~(\ref{eq:twoFreqDrivingForce})) oscillating at
resonance frequency, $\omega_1=\omega_0$. Furthermore, we concentrate on the
effect of a dissipative surface force of the form
(\ref{eq:PlatzDissip}). The integral then reduces to
Eq.~(\ref{eq:ptResult}). The implications of this result are discussed
in the main text.

\section{Commutators of functions of position and momentum}\label{sec:commutator}
We consider functions of operator
$ \hat{x}$ and want to evaluate commutators of
the form 
\begin{align} 
   X_n &= [V(\hat{x}), \hat{p}^n ] .
\end{align} 
We note that, since the commutator between $\delta\hat{x}$ and
$\delta\hat{p}$ of Eqs.~(\ref{eq:defDeltaX}) and (\ref{eq:defDeltaP})
is the same as that of $\hat{x}$ and $\hat{p}$, our
results are also valid for commutators of the form
$ [V(\delta\hat{x}), \delta\hat{p}^n ] $.

{\bf Lemma 1: } 
\begin{align} 
  X_n &= i \sum_{l=1}^{n} c_{n,l} \hbar^l R^{(n-l)}(V^{(l)})
\label{eq:XnForm}\\
  R^{(m)}(f(\hat{x})) &= \hat{p}^m f(\hat{x})
   +f(\hat{x})\hat{p}^m,
\end{align} 
with coefficients $c_{n,l}$ that need to be determined.

{\em Proof:} For $n=1$ we have 
\begin{align} 
  X_1 &=  i\hbar c_{1,1} R^{(0)}(V^{(1)}),
\end{align} 
with $c_{1,1}=\frac{ 1}{2}$. Assuming relation (\ref{eq:XnForm}) holds
for $n-1$, we obtain
\begin{align} 
  X_n &=
   i \left (
  \frac{\hbar}{2} R^{(n-1)}(V') + 
  \frac{ 1}{2} \sum_{r=1}^{n-1} c_{n-1,r} \hbar^r \left (  \hat{p} R^{(n-1-r)}(V^{(r)})
    +  R^{(n-1-r)}(V^{(r)})\hat{p} \right )
  \right ).
\end{align} 
Now,
\begin{align} 
\hat{p} \, R^{(m)}(f)
    +  R^{(m)}(f)\hat{p} &=
  2R^{(m+1)}(f) 
  +
  \sum_{k=1}^{m} c_{m,k} \hbar^{k+1}R^{(m-k)}(f^{(k+1)}),
\end{align} 
so that,
\begin{align} 
  X_n &= i \left (
  \frac{\hbar}{2} R^{(n-1)}(V') + 
  \sum_{l=1}^{n-1} c_{n-1,l} \hbar^l
   R^{(n-l)}(V^{(l)})
  +\frac{1 }{2}
  \sum_{r=1}^{n-2} c_{n-1,r} 
  \sum_{k=1}^{n-1-r} c_{n-1-r,k}
  \hbar^{r+k+1} R^{(n-1-r-k)}(V^{(r+k+1)})
  \right ).
\end{align} 
Introducing the new summation index $l=r+k+1$, we find
\begin{align} 
  X_n  &= i \left (
  \frac{\hbar}{2} R^{(n-1)}(V') + 
  \sum_{l=1}^{n-1} c_{n-1,l} \hbar^l
   R^{(n-l)}(V^{(l)})
  +\frac{1}{2}
  \sum_{l=3}^{n}  \hbar^l R^{(n-l)}(V^{(l)}) \sum_{r=1}^{l-2} c_{n-1,r} 
  c_{n-1-r,l-r-1} 
  \right )
\\ &=
   i \sum_{l=1}^{n} c_{n,l} \hbar^l
   R^{(n-l)}(V^{(l)}),
\end{align} 
with
\begin{align} 
   c_{n,1} &=\frac{ 1}{2} + c_{n-1,1}
\\
  c_{n,2} &= c_{n-1,2}
\\ 
   c_{n,l} &= c_{n-1,l} +\frac{1}{2}
   \sum_{r=1}^{l-2} c_{n-1,r} 
  c_{n-1-r,l-r-1}
  \quad 
  \text{ for }  3 \leq l \leq n.
\label{eq:CnRecursion}\end{align} 
This completes the proof of lemma 1.

Eq.~(\ref{eq:CnRecursion}) provides us with a recursion relation that
can be used to determine all factors $c_{n,l}$. We have verified that,
up to $n=15$, these factors correspond to coefficients of Euler
polynomials $E_n(x)$. More specifically, we found that
\begin{align} 
  \sum_{l=1}^n c_{n,l} x^{n-l} &=
  i\left (
  i^n E_n(-ix) -x^n
  \right ).
\end{align} 

We can now introduce super-operators defined by
\begin{align} 
   \overset{\rightarrow}{{\cal P}}^m V(\hat{x}) &=
    \left ( \frac{\hat{p}^m}{\hbar
    \frac{ \partial }{\partial \hat{x}}}
  \right )^m V(\hat{x}) 
\\
   \overset{\leftarrow}{{\cal P}}^m V(\hat{x}) &=
    \left ( \frac{1}{\hbar
    \frac{ \partial }{\partial \hat{x}}}
  \right )^m V(\hat{x}) \hat{p}^m ,
\end{align} 
to write the commutation relations in a compact form,
\begin{align} 
   [V(\hat{x}), \hat{p}^n ] &= 
  - \left ( \hbar \frac{ \partial }{\partial \hat{x}}
   \right )^n \left (
   i^n E_n (-i \overset{\rightarrow}{{\cal P}})
   - \overset{\rightarrow}{{\cal P}}^n 
  + i^n E_n (-i \overset{\leftarrow}{{\cal P}})
   - \overset{\leftarrow}{{\cal P}}^n 
  \right )  V(\hat{x}).
\label{eq:VpnCommutator}\end{align} 
It may appear strange that a derivative operator
appears in the denominator of super-operators 
$\overset{\leftrightarrow}{{\cal P}}$. However, no negative powers of
derivative operators appear in result (\ref{eq:VpnCommutator}).
We remark that a similar result for quadratic potentials has been
proven by De Angelis and Vignat \cite{DeAngelis2015}.

An important special case is when the function is a power, $V(
\hat{x}) =\hat{x}^m$. It is easy to see that the (mean value of
the) commutator then
reduces to
\begin{align} 
  \langle [ \delta\hat{x}^m, \delta\hat{p}^n] \rangle &=
  2i \sum_{l=1}^{\text{min}(m,n)} \binom{m}{l} 
  \frac{c_{n,l}\hbar^l}{l!} 
  \Delta_{m-l,n-l}.
\label{eq:xPowerpPowerComm}\end{align}

\end{appendix}

\end{widetext}

\bibliographystyle{apsrev4-1}
\bibliography{/Users/pmarzlin/Documents/literatur/kpmJabRef}

\begin{thebibliography}{41}%
\makeatletter
\providecommand \@ifxundefined [1]{%
 \@ifx{#1\undefined}
}%
\providecommand \@ifnum [1]{%
 \ifnum #1\expandafter \@firstoftwo
 \else \expandafter \@secondoftwo
 \fi
}%
\providecommand \@ifx [1]{%
 \ifx #1\expandafter \@firstoftwo
 \else \expandafter \@secondoftwo
 \fi
}%
\providecommand \natexlab [1]{#1}%
\providecommand \enquote  [1]{``#1''}%
\providecommand \bibnamefont  [1]{#1}%
\providecommand \bibfnamefont [1]{#1}%
\providecommand \citenamefont [1]{#1}%
\providecommand \href@noop [0]{\@secondoftwo}%
\providecommand \href [0]{\begingroup \@sanitize@url \@href}%
\providecommand \@href[1]{\@@startlink{#1}\@@href}%
\providecommand \@@href[1]{\endgroup#1\@@endlink}%
\providecommand \@sanitize@url [0]{\catcode `\\12\catcode `\$12\catcode
  `\&12\catcode `\#12\catcode `\^12\catcode `\_12\catcode `\%12\relax}%
\providecommand \@@startlink[1]{}%
\providecommand \@@endlink[0]{}%
\providecommand \url  [0]{\begingroup\@sanitize@url \@url }%
\providecommand \@url [1]{\endgroup\@href {#1}{\urlprefix }}%
\providecommand \urlprefix  [0]{URL }%
\providecommand \Eprint [0]{\href }%
\providecommand \doibase [0]{http://dx.doi.org/}%
\providecommand \selectlanguage [0]{\@gobble}%
\providecommand \bibinfo  [0]{\@secondoftwo}%
\providecommand \bibfield  [0]{\@secondoftwo}%
\providecommand \translation [1]{[#1]}%
\providecommand \BibitemOpen [0]{}%
\providecommand \bibitemStop [0]{}%
\providecommand \bibitemNoStop [0]{.\EOS\space}%
\providecommand \EOS [0]{\spacefactor3000\relax}%
\providecommand \BibitemShut  [1]{\csname bibitem#1\endcsname}%
\let\auto@bib@innerbib\@empty
\bibitem [{\citenamefont {L\"uth}(2015)}]{Lueth2015}%
  \BibitemOpen
  \bibfield  {author} {\bibinfo {author} {\bibfnamefont {H.}~\bibnamefont
  {L\"uth}},\ }\href {\doibase https://doi.org/10.1007/978-3-319-10756-1_1}
  {\emph {\bibinfo {title} {Solid Surfaces, Interfaces and Thin Films}}}\
  (\bibinfo  {publisher} {Springer},\ \bibinfo {year} {2015})\BibitemShut
  {NoStop}%
\bibitem [{\citenamefont {Binnig}\ \emph {et~al.}(1986)\citenamefont {Binnig},
  \citenamefont {Quate},\ and\ \citenamefont {Gerber}}]{PhysRevLett.56.930}%
  \BibitemOpen
  \bibfield  {author} {\bibinfo {author} {\bibfnamefont {G.}~\bibnamefont
  {Binnig}}, \bibinfo {author} {\bibfnamefont {C.~F.}\ \bibnamefont {Quate}}, \
  and\ \bibinfo {author} {\bibfnamefont {C.}~\bibnamefont {Gerber}},\ }\href
  {\doibase 10.1103/PhysRevLett.56.930} {\bibfield  {journal} {\bibinfo
  {journal} {Phys. Rev. Lett.}\ }\textbf {\bibinfo {volume} {56}},\ \bibinfo
  {pages} {930} (\bibinfo {year} {1986})}\BibitemShut {NoStop}%
\bibitem [{\citenamefont {Gavara}(2017)}]{Gavara2017}%
  \BibitemOpen
  \bibfield  {author} {\bibinfo {author} {\bibfnamefont {N.}~\bibnamefont
  {Gavara}},\ }\href {\doibase https://doi.org/10.1002/jemt.22776} {\bibfield
  {journal} {\bibinfo  {journal} {Microsc. Res. Tech.}\ }\textbf {\bibinfo
  {volume} {80}},\ \bibinfo {pages} {75} (\bibinfo {year} {2017})}\BibitemShut
  {NoStop}%
\bibitem [{\citenamefont {Magonov}\ and\ \citenamefont
  {Whangbo}(2008)}]{Magonov2008}%
  \BibitemOpen
  \bibfield  {author} {\bibinfo {author} {\bibfnamefont {S.~N.}\ \bibnamefont
  {Magonov}}\ and\ \bibinfo {author} {\bibfnamefont {M.}~\bibnamefont
  {Whangbo}},\ }\href@noop {} {\emph {\bibinfo {title} {Surface analysis with
  STM and AFM: experimental and theoretical aspects of image analysis}}}\
  (\bibinfo  {publisher} {John Wiley and Sons},\ \bibinfo {year}
  {2008})\BibitemShut {NoStop}%
\bibitem [{\citenamefont {Hinterdorfer}\ and\ \citenamefont
  {Dufr\^ene}(2006)}]{Hinterdorfer2006}%
  \BibitemOpen
  \bibfield  {author} {\bibinfo {author} {\bibfnamefont {P.}~\bibnamefont
  {Hinterdorfer}}\ and\ \bibinfo {author} {\bibfnamefont {Y.}~\bibnamefont
  {Dufr\^ene}},\ }\href {\doibase 10.1038/nmeth871} {\bibfield  {journal}
  {\bibinfo  {journal} {Nature Methods}\ }\textbf {\bibinfo {volume} {3}},\
  \bibinfo {pages} {347} (\bibinfo {year} {2006})}\BibitemShut {NoStop}%
\bibitem [{\citenamefont {Quigley}\ \emph {et~al.}(2016)\citenamefont
  {Quigley}, \citenamefont {Veres},\ and\ \citenamefont
  {Kreplak}}]{Kreplak2016}%
  \BibitemOpen
  \bibfield  {author} {\bibinfo {author} {\bibfnamefont {A.}~\bibnamefont
  {Quigley}}, \bibinfo {author} {\bibfnamefont {S.~P.}\ \bibnamefont {Veres}},
  \ and\ \bibinfo {author} {\bibfnamefont {L.}~\bibnamefont {Kreplak}},\ }\href
  {\doibase 10.1371/journal.pone.0161951} {\bibfield  {journal} {\bibinfo
  {journal} {PLoS One}\ }\textbf {\bibinfo {volume} {11}},\ \bibinfo {pages}
  {e0161951} (\bibinfo {year} {2016})}\BibitemShut {NoStop}%
\bibitem [{\citenamefont {Pyrgiotakis}\ \emph {et~al.}(2014)\citenamefont
  {Pyrgiotakis}, \citenamefont {Blattmann},\ and\ \citenamefont
  {Demokritou}}]{Pyrgiotakis2014}%
  \BibitemOpen
  \bibfield  {author} {\bibinfo {author} {\bibfnamefont {G.}~\bibnamefont
  {Pyrgiotakis}}, \bibinfo {author} {\bibfnamefont {C.~O.}\ \bibnamefont
  {Blattmann}}, \ and\ \bibinfo {author} {\bibfnamefont {P.}~\bibnamefont
  {Demokritou}},\ }\href {\doibase 10.1021/sc500152g} {\bibfield  {journal}
  {\bibinfo  {journal} {ACS Sustainable Chem. Eng.}\ }\textbf {\bibinfo
  {volume} {2}},\ \bibinfo {pages} {1681} (\bibinfo {year} {2014})}\BibitemShut
  {NoStop}%
\bibitem [{\citenamefont {Garc\'ia}\ and\ \citenamefont
  {P\'erez}(2002)}]{Garcia2002}%
  \BibitemOpen
  \bibfield  {author} {\bibinfo {author} {\bibfnamefont {R.}~\bibnamefont
  {Garc\'ia}}\ and\ \bibinfo {author} {\bibfnamefont {R.}~\bibnamefont
  {P\'erez}},\ }\href {\doibase https://doi.org/10.1016/S0167-5729(02)00077-8}
  {\bibfield  {journal} {\bibinfo  {journal} {Surf. Sci. Rep.}\ }\textbf
  {\bibinfo {volume} {47}},\ \bibinfo {pages} {197} (\bibinfo {year}
  {2002})}\BibitemShut {NoStop}%
\bibitem [{\citenamefont {Platz}\ \emph {et~al.}(2008)\citenamefont {Platz},
  \citenamefont {Thol\'en}, \citenamefont {Pesen},\ and\ \citenamefont
  {Haviland}}]{Platz2008}%
  \BibitemOpen
  \bibfield  {author} {\bibinfo {author} {\bibfnamefont {D.}~\bibnamefont
  {Platz}}, \bibinfo {author} {\bibfnamefont {E.}~\bibnamefont {Thol\'en}},
  \bibinfo {author} {\bibfnamefont {D.}~\bibnamefont {Pesen}}, \ and\ \bibinfo
  {author} {\bibfnamefont {D.}~\bibnamefont {Haviland}},\ }\href {\doibase
  10.1063/1.2909569} {\bibfield  {journal} {\bibinfo  {journal} {Appl. Phys.
  Lett.}\ }\textbf {\bibinfo {volume} {92}},\ \bibinfo {pages} {153106}
  (\bibinfo {year} {2008})}\BibitemShut {NoStop}%
\bibitem [{\citenamefont {Passian}\ and\ \citenamefont
  {Siopsis}(2016)}]{PhysRevA.94.023812}%
  \BibitemOpen
  \bibfield  {author} {\bibinfo {author} {\bibfnamefont {A.}~\bibnamefont
  {Passian}}\ and\ \bibinfo {author} {\bibfnamefont {G.}~\bibnamefont
  {Siopsis}},\ }\href {\doibase 10.1103/PhysRevA.94.023812} {\bibfield
  {journal} {\bibinfo  {journal} {Phys. Rev. A}\ }\textbf {\bibinfo {volume}
  {94}},\ \bibinfo {pages} {023812} (\bibinfo {year} {2016})}\BibitemShut
  {NoStop}%
\bibitem [{\citenamefont {Passian}\ and\ \citenamefont
  {Siopsis}(2017)}]{Passian2017}%
  \BibitemOpen
  \bibfield  {author} {\bibinfo {author} {\bibfnamefont {A.}~\bibnamefont
  {Passian}}\ and\ \bibinfo {author} {\bibfnamefont {G.}~\bibnamefont
  {Siopsis}},\ }\href {\doibase 10.1103/PhysRevA.95.043812} {\bibfield
  {journal} {\bibinfo  {journal} {Phys. Rev. A}\ }\textbf {\bibinfo {volume}
  {95}},\ \bibinfo {pages} {043812} (\bibinfo {year} {2017})}\BibitemShut
  {NoStop}%
\bibitem [{\citenamefont {Mandel}\ and\ \citenamefont
  {Wolf}(1995)}]{MandelWolf}%
  \BibitemOpen
  \bibfield  {author} {\bibinfo {author} {\bibfnamefont {L.}~\bibnamefont
  {Mandel}}\ and\ \bibinfo {author} {\bibfnamefont {E.}~\bibnamefont {Wolf}},\
  }\href@noop {} {\emph {\bibinfo {title} {Optical Coherence and Quantum
  Optics}}}\ (\bibinfo  {publisher} {Cambridge University Press},\ \bibinfo
  {address} {New York},\ \bibinfo {year} {1995})\BibitemShut {NoStop}%
\bibitem [{\citenamefont {Aspelmeyer}\ \emph {et~al.}(2014)\citenamefont
  {Aspelmeyer}, \citenamefont {Kippenberg},\ and\ \citenamefont
  {Marquardt}}]{Aspelmeyer2014Book}%
  \BibitemOpen
  \bibinfo {editor} {\bibfnamefont {M.}~\bibnamefont {Aspelmeyer}}, \bibinfo
  {editor} {\bibfnamefont {T.}~\bibnamefont {Kippenberg}}, \ and\ \bibinfo
  {editor} {\bibfnamefont {F.}~\bibnamefont {Marquardt}},\ eds.,\ \href@noop {}
  {\emph {\bibinfo {title} {Cavity Optomechanics: Nano- and Micromechanical
  Resonators Interacting with Light}}}\ (\bibinfo  {publisher} {Springer},\
  \bibinfo {year} {2014})\BibitemShut {NoStop}%
\bibitem [{\citenamefont {Bowen}\ and\ \citenamefont
  {Milburn}(2016)}]{Bowen2016Book}%
  \BibitemOpen
  \bibfield  {author} {\bibinfo {author} {\bibfnamefont {W.}~\bibnamefont
  {Bowen}}\ and\ \bibinfo {author} {\bibfnamefont {G.}~\bibnamefont
  {Milburn}},\ }\href@noop {} {\emph {\bibinfo {title} {Quantum
  Optomechanics}}}\ (\bibinfo  {publisher} {CRC Press},\ \bibinfo {year}
  {2016})\BibitemShut {NoStop}%
\bibitem [{\citenamefont {Riedinger}\ \emph {et~al.}(2018)\citenamefont
  {Riedinger}, \citenamefont {Wallucks}, \citenamefont {Marinkovi\'c},
  \citenamefont {L\"oschnauer}, \citenamefont {Aspelmeyer}, \citenamefont
  {Hong},\ and\ \citenamefont {Gr\"oblacher}}]{Riedinger2018}%
  \BibitemOpen
  \bibfield  {author} {\bibinfo {author} {\bibfnamefont {R.}~\bibnamefont
  {Riedinger}}, \bibinfo {author} {\bibfnamefont {A.}~\bibnamefont {Wallucks}},
  \bibinfo {author} {\bibfnamefont {I.}~\bibnamefont {Marinkovi\'c}}, \bibinfo
  {author} {\bibfnamefont {C.}~\bibnamefont {L\"oschnauer}}, \bibinfo {author}
  {\bibfnamefont {M.}~\bibnamefont {Aspelmeyer}}, \bibinfo {author}
  {\bibfnamefont {S.}~\bibnamefont {Hong}}, \ and\ \bibinfo {author}
  {\bibfnamefont {S.}~\bibnamefont {Gr\"oblacher}},\ }\href
  {https://doi.org/10.1038/s41586-018-0036-z} {\bibfield  {journal} {\bibinfo
  {journal} {Nature}\ }\textbf {\bibinfo {volume} {556}},\ \bibinfo {pages}
  {473} (\bibinfo {year} {2018})}\BibitemShut {NoStop}%
\bibitem [{\citenamefont {Mercier~de L{\'e}pinay}\ \emph
  {et~al.}(2021)\citenamefont {Mercier~de L{\'e}pinay}, \citenamefont
  {Ockeloen-Korppi}, \citenamefont {Woolley},\ and\ \citenamefont
  {Sillanp{\"a}{\"a}}}]{MercierdeLepinay2021}%
  \BibitemOpen
  \bibfield  {author} {\bibinfo {author} {\bibfnamefont {L.}~\bibnamefont
  {Mercier~de L{\'e}pinay}}, \bibinfo {author} {\bibfnamefont {C.~F.}\
  \bibnamefont {Ockeloen-Korppi}}, \bibinfo {author} {\bibfnamefont {M.~J.}\
  \bibnamefont {Woolley}}, \ and\ \bibinfo {author} {\bibfnamefont {M.~A.}\
  \bibnamefont {Sillanp{\"a}{\"a}}},\ }\href {\doibase 10.1126/science.abf5389}
  {\bibfield  {journal} {\bibinfo  {journal} {Science}\ }\textbf {\bibinfo
  {volume} {372}},\ \bibinfo {pages} {625} (\bibinfo {year}
  {2021})}\BibitemShut {NoStop}%
\bibitem [{\citenamefont {Yu}\ \emph {et~al.}(2020)\citenamefont {Yu} \emph
  {et~al.}}]{Yu2020}%
  \BibitemOpen
  \bibfield  {author} {\bibinfo {author} {\bibfnamefont {H.}~\bibnamefont {Yu}}
  \emph {et~al.},\ }\href {https://doi.org/10.1038/s41586-020-2420-8}
  {\bibfield  {journal} {\bibinfo  {journal} {Nature}\ }\textbf {\bibinfo
  {volume} {583}},\ \bibinfo {pages} {43} (\bibinfo {year} {2020})}\BibitemShut
  {NoStop}%
\bibitem [{\citenamefont {Gardiner}\ and\ \citenamefont
  {Zoller}(2004)}]{GardinerZollerQuantumNoise}%
  \BibitemOpen
  \bibfield  {author} {\bibinfo {author} {\bibfnamefont {C.~W.}\ \bibnamefont
  {Gardiner}}\ and\ \bibinfo {author} {\bibfnamefont {P.}~\bibnamefont
  {Zoller}},\ }\href
  {http://books.google.ca/books?id=a_xsT8oGhdgC&lpg=PP1&ots=kXx1vVbWu9&dq=gardiner%20zoller&pg=PR4#v=onepage&q&f=false}
  {\emph {\bibinfo {title} {Quantum Noise}}}\ (\bibinfo  {publisher} {Springer,
  Berlin},\ \bibinfo {year} {2004})\BibitemShut {NoStop}%
\bibitem [{\citenamefont {Benaglia}\ \emph {et~al.}(2019)\citenamefont
  {Benaglia}, \citenamefont {Amo},\ and\ \citenamefont
  {Garcia}}]{Benaglia2019}%
  \BibitemOpen
  \bibfield  {author} {\bibinfo {author} {\bibfnamefont {S.}~\bibnamefont
  {Benaglia}}, \bibinfo {author} {\bibfnamefont {C.}~\bibnamefont {Amo}}, \
  and\ \bibinfo {author} {\bibfnamefont {R.}~\bibnamefont {Garcia}},\ }\href
  {\doibase 10.1039/C9NR04396A} {\bibfield  {journal} {\bibinfo  {journal}
  {Nanoscale}\ }\textbf {\bibinfo {volume} {11}},\ \bibinfo {pages} {15289}
  (\bibinfo {year} {2019})}\BibitemShut {NoStop}%
\bibitem [{\citenamefont {Johnson}\ \emph {et~al.}(1971)\citenamefont
  {Johnson}, \citenamefont {Kendall}, \citenamefont {Roberts},\ and\
  \citenamefont {Tabor}}]{Johnson1971}%
  \BibitemOpen
  \bibfield  {author} {\bibinfo {author} {\bibfnamefont {K.}~\bibnamefont
  {Johnson}}, \bibinfo {author} {\bibfnamefont {K.}~\bibnamefont {Kendall}},
  \bibinfo {author} {\bibfnamefont {A.}~\bibnamefont {Roberts}}, \ and\
  \bibinfo {author} {\bibfnamefont {D.}~\bibnamefont {Tabor}},\ }\href
  {\doibase 10.1098/rspa.1971.0141} {\bibfield  {journal} {\bibinfo  {journal}
  {Proc R Soc Lond A}\ }\textbf {\bibinfo {volume} {324}},\ \bibinfo {pages}
  {301} (\bibinfo {year} {1971})}\BibitemShut {NoStop}%
\bibitem [{Dan(2008)}]{Dankowicz2008}%
  \BibitemOpen
  \href {\doibase 10.1115/ESDA2008-59293} {\emph {\bibinfo {title}
  {Discontinuity-Induced Bifurcations in Systems With Hysteretic Force
  Interactions}}},\ \bibinfo {series} {Engineering Systems Design and
  Analysis}, Vol.\ \bibinfo {volume} {Volume 2: Automotive Systems;
  Bioengineering and Biomedical Technology; Computational Mechanics; Controls;
  Dynamical Systems}\ (\bibinfo {year} {2008})\BibitemShut {NoStop}%
\bibitem [{\citenamefont {Lee}\ and\ \citenamefont {Radok}(1960)}]{Lee1960}%
  \BibitemOpen
  \bibfield  {author} {\bibinfo {author} {\bibfnamefont {E.~H.}\ \bibnamefont
  {Lee}}\ and\ \bibinfo {author} {\bibfnamefont {J.~R.~M.}\ \bibnamefont
  {Radok}},\ }\href {\doibase 10.1115/1.3644020} {\bibfield  {journal}
  {\bibinfo  {journal} {J. Appl. Mech.}\ }\textbf {\bibinfo {volume} {27}},\
  \bibinfo {pages} {438} (\bibinfo {year} {1960})}\BibitemShut {NoStop}%
\bibitem [{\citenamefont {Ting}(1966)}]{Ting1966}%
  \BibitemOpen
  \bibfield  {author} {\bibinfo {author} {\bibfnamefont {T.~C.~T.}\
  \bibnamefont {Ting}},\ }\href {\doibase 10.1115/1.3625192} {\bibfield
  {journal} {\bibinfo  {journal} {J. Appl. Mech.}\ }\textbf {\bibinfo {volume}
  {33}},\ \bibinfo {pages} {845} (\bibinfo {year} {1966})}\BibitemShut
  {NoStop}%
\bibitem [{\citenamefont {Platz}\ \emph {et~al.}(2013)\citenamefont {Platz},
  \citenamefont {Forchheimer}, \citenamefont {Tholén},\ and\ \citenamefont
  {Haviland}}]{Platz2013}%
  \BibitemOpen
  \bibfield  {author} {\bibinfo {author} {\bibfnamefont {D.}~\bibnamefont
  {Platz}}, \bibinfo {author} {\bibfnamefont {D.}~\bibnamefont {Forchheimer}},
  \bibinfo {author} {\bibfnamefont {E.~A.}\ \bibnamefont {Tholén}}, \ and\
  \bibinfo {author} {\bibfnamefont {D.~B.}\ \bibnamefont {Haviland}},\ }\href
  {https://doi.org/10.1038/ncomms2365} {\bibfield  {journal} {\bibinfo
  {journal} {Nature Communications}\ }\textbf {\bibinfo {volume} {4}},\
  \bibinfo {pages} {1360} (\bibinfo {year} {2013})}\BibitemShut {NoStop}%
\bibitem [{\citenamefont {Spanos}(1999)}]{SpanosProbability}%
  \BibitemOpen
  \bibfield  {author} {\bibinfo {author} {\bibfnamefont {A.}~\bibnamefont
  {Spanos}},\ }\href@noop {} {\emph {\bibinfo {title} {Probability Theory and
  Statistical Inference: Econometric Modeling with Observational Data}}}\
  (\bibinfo  {publisher} {Cambridge University Press},\ \bibinfo {year}
  {1999})\BibitemShut {NoStop}%
\bibitem [{\citenamefont {Liouville}(1838)}]{Liouville1838}%
  \BibitemOpen
  \bibfield  {author} {\bibinfo {author} {\bibfnamefont {J.}~\bibnamefont
  {Liouville}},\ }\href
  {http://sites.mathdoc.fr/JMPA/PDF/JMPA_1838_1_3_A26_0.pdf} {\bibfield
  {journal} {\bibinfo  {journal} {J. Math. Pures Appl.}\ }\textbf {\bibinfo
  {volume} {3}},\ \bibinfo {pages} {342–349} (\bibinfo {year}
  {1838})}\BibitemShut {NoStop}%
\bibitem [{\citenamefont {Streater}\ and\ \citenamefont
  {Wightman}(1964)}]{StreaterWightman}%
  \BibitemOpen
  \bibfield  {author} {\bibinfo {author} {\bibfnamefont {R.~F.}\ \bibnamefont
  {Streater}}\ and\ \bibinfo {author} {\bibfnamefont {A.~S.}\ \bibnamefont
  {Wightman}},\ }\href@noop {} {\emph {\bibinfo {title} {PCT, Spin and
  Statistics and All That}}}\ (\bibinfo  {publisher} {Princeton University
  Press},\ \bibinfo {year} {1964})\BibitemShut {NoStop}%
\bibitem [{\citenamefont {Azzalini}\ and\ \citenamefont
  {Valle}(1996)}]{Azzalini1996}%
  \BibitemOpen
  \bibfield  {author} {\bibinfo {author} {\bibfnamefont {A.}~\bibnamefont
  {Azzalini}}\ and\ \bibinfo {author} {\bibfnamefont {A.~D.}\ \bibnamefont
  {Valle}},\ }\href {https://www.jstor.org/stable/2337278} {\bibfield
  {journal} {\bibinfo  {journal} {Biometrika}\ }\textbf {\bibinfo {volume}
  {83}},\ \bibinfo {pages} {715} (\bibinfo {year} {1996})}\BibitemShut
  {NoStop}%
\bibitem [{\citenamefont {Callen}\ and\ \citenamefont
  {Welton}(1951)}]{PhysRev.83.34}%
  \BibitemOpen
  \bibfield  {author} {\bibinfo {author} {\bibfnamefont {H.}~\bibnamefont
  {Callen}}\ and\ \bibinfo {author} {\bibfnamefont {T.}~\bibnamefont
  {Welton}},\ }\href {\doibase 10.1103/PhysRev.83.34} {\bibfield  {journal}
  {\bibinfo  {journal} {Phys. Rev.}\ }\textbf {\bibinfo {volume} {83}},\
  \bibinfo {pages} {34} (\bibinfo {year} {1951})}\BibitemShut {NoStop}%
\bibitem [{\citenamefont {Kubo}(1966)}]{Kubo1966}%
  \BibitemOpen
  \bibfield  {author} {\bibinfo {author} {\bibfnamefont {R.}~\bibnamefont
  {Kubo}},\ }\href {\doibase 10.1088/0034-4885/29/1/306} {\bibfield  {journal}
  {\bibinfo  {journal} {Rep. Prog. Phys.}\ }\textbf {\bibinfo {volume} {29}},\
  \bibinfo {pages} {255} (\bibinfo {year} {1966})}\BibitemShut {NoStop}%
\bibitem [{\citenamefont {Hahn}(1950)}]{PhysRev.80.580}%
  \BibitemOpen
  \bibfield  {author} {\bibinfo {author} {\bibfnamefont {E.~L.}\ \bibnamefont
  {Hahn}},\ }\href {\doibase 10.1103/PhysRev.80.580} {\bibfield  {journal}
  {\bibinfo  {journal} {Phys. Rev.}\ }\textbf {\bibinfo {volume} {80}},\
  \bibinfo {pages} {580} (\bibinfo {year} {1950})}\BibitemShut {NoStop}%
\bibitem [{\citenamefont {H\"ansch}\ \emph {et~al.}(1975)\citenamefont
  {H\"ansch}, \citenamefont {Lee}, \citenamefont {Wallenstein},\ and\
  \citenamefont {Wieman}}]{PhysRevLett.34.307}%
  \BibitemOpen
  \bibfield  {author} {\bibinfo {author} {\bibfnamefont {T.~W.}\ \bibnamefont
  {H\"ansch}}, \bibinfo {author} {\bibfnamefont {S.~A.}\ \bibnamefont {Lee}},
  \bibinfo {author} {\bibfnamefont {R.}~\bibnamefont {Wallenstein}}, \ and\
  \bibinfo {author} {\bibfnamefont {C.}~\bibnamefont {Wieman}},\ }\href
  {\doibase 10.1103/PhysRevLett.34.307} {\bibfield  {journal} {\bibinfo
  {journal} {Phys. Rev. Lett.}\ }\textbf {\bibinfo {volume} {34}},\ \bibinfo
  {pages} {307} (\bibinfo {year} {1975})}\BibitemShut {NoStop}%
\bibitem [{\citenamefont {Marzlin}\ and\ \citenamefont
  {Audretsch}(1996)}]{PRA53:312}%
  \BibitemOpen
  \bibfield  {author} {\bibinfo {author} {\bibfnamefont {K.-P.}\ \bibnamefont
  {Marzlin}}\ and\ \bibinfo {author} {\bibfnamefont {J.}~\bibnamefont
  {Audretsch}},\ }\href {http://link.aps.org/abstract/PRA/v53/p312} {\bibfield
  {journal} {\bibinfo  {journal} {Phys. Rev. A}\ }\textbf {\bibinfo {volume}
  {53}},\ \bibinfo {pages} {312} (\bibinfo {year} {1996})}\BibitemShut
  {NoStop}%
\bibitem [{\citenamefont {Pottier}\ and\ \citenamefont
  {Bellon}(2017)}]{Pottier2017}%
  \BibitemOpen
  \bibfield  {author} {\bibinfo {author} {\bibfnamefont {B.}~\bibnamefont
  {Pottier}}\ and\ \bibinfo {author} {\bibfnamefont {L.}~\bibnamefont
  {Bellon}},\ }\href {\doibase 10.1063/1.4977790} {\bibfield  {journal}
  {\bibinfo  {journal} {Appl. Phys. Lett.}\ }\textbf {\bibinfo {volume}
  {110}},\ \bibinfo {pages} {094105} (\bibinfo {year} {2017})}\BibitemShut
  {NoStop}%
\bibitem [{\citenamefont {Cercignani}\ \emph {et~al.}(1997)\citenamefont
  {Cercignani}, \citenamefont {Gerasimenko},\ and\ \citenamefont
  {Petrina}}]{Cercignani97}%
  \BibitemOpen
  \bibfield  {author} {\bibinfo {author} {\bibfnamefont {C.}~\bibnamefont
  {Cercignani}}, \bibinfo {author} {\bibfnamefont {V.}~\bibnamefont
  {Gerasimenko}}, \ and\ \bibinfo {author} {\bibfnamefont {D.}~\bibnamefont
  {Petrina}},\ }\href {\doibase 10.1007/978-94-011-5558-8_2} {\emph {\bibinfo
  {title} {Many-Particle Dynamics and Kinetic Equations}}}\ (\bibinfo
  {publisher} {Springer Nature},\ \bibinfo {year} {1997})\BibitemShut {NoStop}%
\bibitem [{\citenamefont {Fokker}(1914)}]{Fokker1914}%
  \BibitemOpen
  \bibfield  {author} {\bibinfo {author} {\bibfnamefont {A.~D.}\ \bibnamefont
  {Fokker}},\ }\href {\doibase https://doi.org/10.1002/andp.19143480507}
  {\bibfield  {journal} {\bibinfo  {journal} {Ann. Phys. (Leipzig)}\ }\textbf
  {\bibinfo {volume} {348}},\ \bibinfo {pages} {810} (\bibinfo {year}
  {1914})}\BibitemShut {NoStop}%
\bibitem [{\citenamefont {Planck}(1917)}]{Planck1917}%
  \BibitemOpen
  \bibfield  {author} {\bibinfo {author} {\bibfnamefont {M.}~\bibnamefont
  {Planck}},\ }\href
  {https://www.biodiversitylibrary.org/page/29213319#page/360/mode/1up}
  {\bibfield  {journal} {\bibinfo  {journal} {Sitzungsber. Preuss. Akad.
  Wiss.}\ }\textbf {\bibinfo {volume} {24}},\ \bibinfo {pages} {324–341}
  (\bibinfo {year} {1917})}\BibitemShut {NoStop}%
\bibitem [{\citenamefont {Lindblad}(1976)}]{linblad:cmp76}%
  \BibitemOpen
  \bibfield  {author} {\bibinfo {author} {\bibfnamefont {G.}~\bibnamefont
  {Lindblad}},\ }\href {\doibase 10.1007/BF01608499} {\bibfield  {journal}
  {\bibinfo  {journal} {Comm. Math. Phys.}\ }\textbf {\bibinfo {volume} {48}},\
  \bibinfo {pages} {119} (\bibinfo {year} {1976})}\BibitemShut {NoStop}%
\bibitem [{\citenamefont {Caldeira}\ and\ \citenamefont
  {Leggett}(1983)}]{Caldeira1983}%
  \BibitemOpen
  \bibfield  {author} {\bibinfo {author} {\bibfnamefont {A.}~\bibnamefont
  {Caldeira}}\ and\ \bibinfo {author} {\bibfnamefont {A.}~\bibnamefont
  {Leggett}},\ }\href {\doibase https://doi.org/10.1016/0003-4916(83)90202-6}
  {\bibfield  {journal} {\bibinfo  {journal} {Ann. Phys.}\ }\textbf {\bibinfo
  {volume} {149}},\ \bibinfo {pages} {374 } (\bibinfo {year}
  {1983})}\BibitemShut {NoStop}%
\bibitem [{\citenamefont {Hornberger}(2009)}]{HornbergerDecoherence}%
  \BibitemOpen
  \bibfield  {author} {\bibinfo {author} {\bibfnamefont {K.}~\bibnamefont
  {Hornberger}},\ }\href {\doibase 10.1007/978-3-540-88169-8} {\bibfield
  {journal} {\bibinfo  {journal} {Lec. Notes Phys.}\ } (\bibinfo {year}
  {2009}),\ 10.1007/978-3-540-88169-8},\ \Eprint
  {http://arxiv.org/abs/quant-ph/0612118v3} {quant-ph/0612118v3} \BibitemShut
  {NoStop}%
\bibitem [{\citenamefont {Angelis}\ and\ \citenamefont
  {Vignat}(2015)}]{DeAngelis2015}%
  \BibitemOpen
  \bibfield  {author} {\bibinfo {author} {\bibfnamefont {V.~D.}\ \bibnamefont
  {Angelis}}\ and\ \bibinfo {author} {\bibfnamefont {C.}~\bibnamefont
  {Vignat}},\ }\href {\doibase 10.1063/1.4938077} {\bibfield  {journal}
  {\bibinfo  {journal} {J. Math. Phys.}\ }\textbf {\bibinfo {volume} {56}},\
  \bibinfo {pages} {123506} (\bibinfo {year} {2015})},\ \Eprint
  {http://arxiv.org/abs/1508.04844} {1508.04844} \BibitemShut {NoStop}%
\end{thebibliography}%

\end{document}